\documentclass[apj]{emulateapj}

\shorttitle{C-COSMOS data analysis}
\shortauthors{Puccetti et al.}

\def\ls{{_<\atop^{\sim}}}

\def\cgs{ ${\rm erg~cm}^{-2}~{\rm s}^{-1}$ }

\begin{document}

\title{ The Chandra survey of the COSMOS field II: source detection and photometry}
\author{S. Puccetti\altaffilmark{1}, C. Vignali\altaffilmark{2,3},  N. Cappelluti\altaffilmark{4},  F. Fiore\altaffilmark{5}, G. Zamorani\altaffilmark{3}, T. L. Aldcroft\altaffilmark{6}, M. Elvis\altaffilmark{6}, R. Gilli\altaffilmark{3} T. Miyaji\altaffilmark{7,8}, H. Brunner\altaffilmark{4}, M. Brusa\altaffilmark{4}, F. Civano\altaffilmark{6}, A. Comastri\altaffilmark{3}, F. Damiani\altaffilmark{9},  A. Fruscione\altaffilmark{4} , A. Finoguenov\altaffilmark{4,10}, A. M. Koekemoer\altaffilmark{11}, V. Mainieri\altaffilmark{12} }

\altaffiltext{1}{ASI Science data Center, via Galileo Galilei,
00044 Frascati, Italy}
\altaffiltext{2}{Dipartimento di Astronomia, Universita' di Bologna,
via Ranzani 1, Bologna, Italy}
\altaffiltext{3}{INAF$-$Osservatorio Astronomico di Bologna, Via
  Ranzani 1, I--40127 Bologna, Italy}
\altaffiltext{4}{Max Planck Institut f\"ur extraterrestrische Physik,
      Giessenbachstrasse 1, D--85748 Garching bei M\"unchen, Germany}
\altaffiltext{5}{INAF-OAR, via Frascati 33, Monteporzio, I00040, Italy}
\altaffiltext{6}{Harvard-Smithsonian Center for Astrophysics, 60
  Garden St., Cambridge, MA 02138 USA}
\altaffiltext{7}{Instituto de Astronom\'ia, Universidad Nacional Aut\'onoma
 de M\'exico, Ensenada, M\'exico (mailing address: PO Box 439027, San Ysidro,
 CA, 92143-9027, USA)}
\altaffiltext{8}{Center for Astrophysics and Space Sciences,
 University of California San Diego, Code 0424, 9500 Gilman Drive,
 La Jolla, CA 92093, USA}
\altaffiltext{9}{INAF - Osservatorio Astronomico di Palermo, Piazza
  del Parlamento 1, I-90134 Palermo, Italy}
\altaffiltext{10}{University of Maryland, Baltimore County, 1000
 Hilltop Circle,  Baltimore, MD 21250, USA}
\altaffiltext{11}{ Space Telescope Science Institute  3700 San Martin Drive, Baltimore MD 21218 U.S.A.}
\altaffiltext{12}{ESO, Karl-Schwarschild-Strasse 2, D--85748 Garching, Germany}

\email{puccetti@asdc.asi.it}
\begin{abstract}

The {\it Chandra} COSMOS Survey (C-COSMOS) is a large, 1.8 Ms, {\it
  Chandra} program, that covers the central contiguous $\sim$~0.92
  deg$^2$ of the COSMOS field. C-COSMOS is the result of a complex
  tiling, with every position being observed in up to six overlapping
  pointings (four overlapping pointings in most of the central
  $\sim$~0.45 deg$^2$ area with the best exposure, and two
  overlapping pointings in most of the surrounding area, covering an
  additional  $\sim$~0.47 deg$^2$). Therefore, the full exploitation of
  the C-COSMOS data requires a dedicated and accurate analysis focused
  on three main issues: 1) maximizing the sensitivity when the PSF
  changes strongly among different observations of the same source
  (from $\sim$~1 arcsec up to $\sim~10$ arcsec half power radius); 2)
  resolving close pairs; and 3) obtaining the best source localization
  and count rate.  We present here our treatment of four key analysis
  items: source detection, localization, photometry, and survey
  sensitivity.  Our final procedure consists of a two step procedure:
  (1) a wavelet detection algorithm, to find source candidates, (2) a
  maximum likelihood Point Spread Function fitting algorithm to evaluate the source
  count rates and the probability that each source candidate is a
  fluctuation of the background.
We discuss the main characteristics of this procedure, that was the result of
detailed comparisons between different detection algorithms and 
photometry tools, calibrated with extensive and dedicated simulations.

\end{abstract}

\section{Introduction}

It is well known that X-ray surveys are an extremely efficient tool to
select Active Galactic Nuclei (AGN). For example in the {\it
XMM-Newton} COSMOS survey, at the 0.5-2 keV limiting flux of
7$\cdot$10$^{-16}$ erg s$^{-1}$ cm$^{-2}$, the AGN surface density is
$\sim$1000 deg$^{-2}$ (Hasinger et al. 2007, Cappelluti et al. 2007),
a factor 2-4 greater than the AGN surface density in the most recent
deep optical surveys, 250 deg$^{-2}$ in the COMBO-17 ( Wolf et
al. 2003) and 470 deg$^{-2}$ in VVDS Survey (Gavignaud et
al. 2006). There are four main causes for the higher efficiency of
X-ray surveys in finding AGN: 1) X-rays directly trace the super
massive black hole (SMBH) accretion, while AGN classification trough
optical line spectroscopy may suffer of uncompleteness and/or
misidentifications; 2) AGN are the dominant X-ray population. In fact
most ($\sim$ 80\%) of the X-ray sources AGN in deep and shallow
surveys turn out to be AGN, unlike at optical wavelengths. 3) 0.5-10
keV X-rays (the typical {\it Chandra} and {\it XMM-Newton} enery band)
are capable to penetrate column densities up to $\sim$10$^{24}$
cm$^{-2}$, allowing the selection of moderately obscured AGN; 4) low
luminosity AGN are difficult to select in optical surveys, because
their light is diluted in the host galaxy emission. 

\begin{figure}
\begin{center}
\includegraphics[angle=0,height=8truecm]{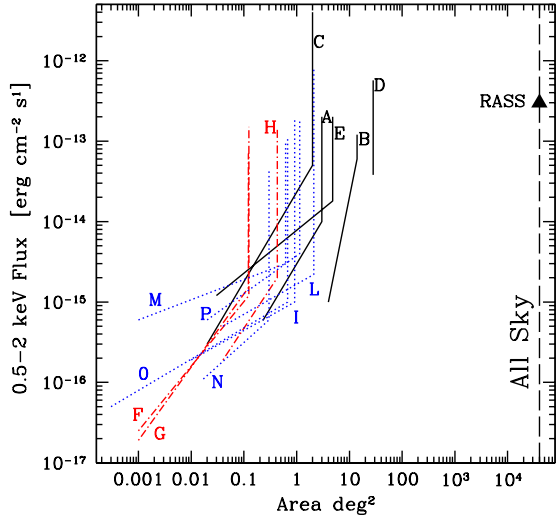}
\caption{The 0.5-2 keV flux range vs. the area coverage for various
surveys. The black solid lines represent few serendipitous surveys:
Helllas2XMM (Baldi et al. 2002, symbol A), CHAMP (Kim et al. 2004a,
2004b, Green et al. 2004, symbol B), SEXSI (Harrison et al. 2003,
symbol C), XMM-BSS (Della Ceca et al. 2004, symbol D), AXIS (Carrera
et al. 2007, symbol E); the red dotted lines represent few deep pencil
beam surveys: CDFN (Brandt et al. 2001, Alexander et al. 2003, symbol
F), CDFS (Giacconi et al. 2001, Luo et al. 2008, symbol G), XMM-Newton
Lockman Hole (Worsley et al. 2004, Brunner et al. 2008, symbol H); the
blue dotted lines represent few wide shallow contiguous surveys:
C-COSMOS (Elvis et al. 2009, symbol I), XMM-COSMOS (Hasinger et
al. 2007, Cappelluti et al. 2007, 2009, symbol L), ELAIS-S1 (Puccetti
et al. 2006, symbol M), E-CDF-S (Lehmer et al. 2005, symbol N),
AEGIS-X (Laird et al. 2009, symbol O), SXDS (Ueda et al. 2008, symbol
P). The black solid triangle represent the ROSAT all sky survey
(RASS, Voges et al. 1999).}
\label{surveys}
\end{center}
\end{figure}

So far {\it Chandra} and {\it XMM-Newton} have performed several deep,
pencil beam, and shallower but wider surveys. Fig. \ref{surveys}
compares the flux limit and area coverage of the main {\it Chandra}
and {\it XMM-Newton} surveys. This figure shows that {\it XMM-Newton}
COSMOS and {\it Chandra}-COSMOS (C-COSMOS, Elvis et al. 2009, Paper I
hereafter) surveys are the deepest surveys on large contiguous area. The
coverage of larger areas at similar flux limits is today achieved only
by serendipitous surveys using mostly not contiguous areas (see e.g.,
CHAMP, Kim et al. 2004a, 2004b, Green et al. 2004).

The Cosmic evolution survey (COSMOS, Scoville et al. 2007) is aimed at
studying the interplay between the Large Scale Structure (LSS) in the
Universe and the formation of galaxies, dark matter, and AGN. The
COSMOS field is located near the equator (10h,+02degrees), covers
$\sim$~2 square degrees as originally defined by the HST/ACS imaging
(Koekemoer et al. 2007), with subsequent deep and extended
multi-wavelength coverage overlapping this area.  The size of COSMOS
was chosen to sample LSS up to a linear size of about 50 Mpc h$^{-1}$
at z~$\sim$~1-2, where AGN and star formation in galaxies are expected
to peak. To study the role of AGN in galaxy evolution the X-ray data
are fundamental. Therefore, the central square degree of the COSMOS
field has been the target of a {\it Chandra} ACIS-I, 1.8~Msec Very
Large Program: the {\it Chandra}-COSMOS survey.

The  C-COSMOS  survey  has  a  rather uniform  effective  exposure  of
$\sim~160$~ksec over a large area ($\sim$~0.45 deg$^2$), thus reaching
$\sim$~3.5 times fainter fluxes than XMM-COSMOS in both 0.5-2 keV band
and  2-7  keV band.  This  flux limit  is  below  the threshold  where
starburst galaxies  become common in X-rays.  The  sharp {\it Chandra}
Point Spread  Function (PSF) allows  nearly unambiguous identification
of optical counterparts (Civano et al. 2009, hereafter Paper III).{\it
Chandra} secures  the identifications of  X-ray sources down  to faint
optical magnitude (i.e.,  I~$\sim$~26), with only $\sim$~2\% ambiguous
identifications,  significantly better  than the  $\sim20\%$ ambiguous
identifications in XMM-Newton (Brusa et al. 2007).

The C-COSMOS survey has a complex tiling (see Fig. \ref{overlay}) in
comparison to other X-ray surveys, in which the overlapping areas of
the single pointings are small and with similar PSFs (see e.g., the
Extended Groth-Streep, AEGIS-X, Laird et al. 2009), or all the
pointings are co-assial and nearly totally overlapping (see e.g.,
CDFS, Giacconi et al. 2001, Luo et al. 2008).  In the C-COSMOS tiling,
the pointings are strongly overlapping and not-coassial. While this
ensures a very uniform sensitivity over most of the field, each source
is observed with up to six different PSFs, requiring the development
of an analysis procedure for data observed with this mixture of
PSF. The procedures presented in this paper are aimed at optimizing
(1) source detection, (2) localization, (3) photometry, and (4) survey
sensitivity. We have made detailed comparisons between different
detection algorithms and photometry tools, testing them extensively on
simulated data. We furthermore validate our results by detailed
inspections of each single source candidate. Our final analysis
consists of a two main steps:

\begin{itemize}

\item [1] \underline{a wavelet detection algorithm}, {\it PWDetect} (Damiani et
al. 1997) is first used to find source candidates. This algorithm is optimized to
cleanly separate nearby sources, to detect point-like sources on top
of extended emission and to give the most accurate positions.

\item [2] \underline{A maximum likelihood PSF fitting algorithm} is then used to
 evaluate the source count rates and the probability that each source
 candidate is not a fluctuation of the background. We used the {\it
 emldetect} algorithm (Cappelluti et al. 2007 and references
 therein). {\it emldetect} works simultaneously with multiple
 overlapping pointings using PSFs appropriate to each one. This
 fitting method ensures accurate evaluation of the survey completeness
 and contamination, efficient deblending and good photometry for close
 pairs, which may be partly blended even at the {\it Chandra}
 resolution.
\end{itemize}

As a third step, we also performed \underline{aperture photometry} for
each candidate X-ray source using 50\%, 90\%, and 95\% encircled count
fractions, using the PSFs appropriate to each observation. The
aperture photometry is also used to check the results. Aperture
photometry is preferable in all cases where the systematic error
introduced by PSF fitting are larger than the statistical errors,
i.e., for bright sources (count rates $\geq 1$ counts/ksec).

The survey sensitivity is limited by both the net (i.e., including
vignetting) exposure time, and by the actual PSF with which a given
region of the area is observed. The latter issue is particularly
relevant for the C-COSMOS tiling. We have developed an algorithm that
evaluates the survey sensitivity at each position on C-COSMOS using a
parameterization of the {\it Chandra} ACIS-I PSF and taking into account
the mixture of PSFs at each position. The resulting sensitivity maps
have been compared and validated with extensive simulations.

The paper is organized as following: in Sect. 2 we briefly present the
C-COSMOS observations and data reduction; we describe the simulations
in Sect. 3; how they were used to select the most efficient detection
algorithm and the final source characterization procedure is
described in Sect. 4; the completeness and reliability are shown in
Sect. 5; in Sect. 6 we apply this procedure to the observed data; in
Sect. 7 we present the calculation of survey sensitivity, the
sky-coverage, and X-ray number counts using the simulated
data. Finally, in Sect. 8 we compare C-COSMOS to a similar {\it
Chandra} survey, i.e., AEGIS-X, and in Sect. 9 we give our
conclusion.

\section{Observations and data reduction}

\begin{figure}
\includegraphics[width=9cm]{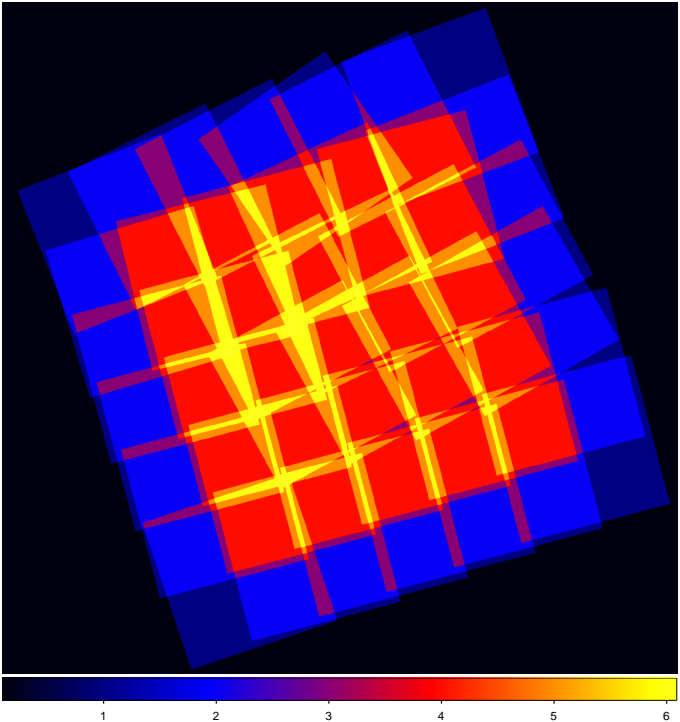}
\caption{The final tiling of the C-COSMOS field, with a color scale
showing the number of the ACIS-I overlapping pointings, as
indicated in the color bar at the bottom of the figure.
\label{overlay}}
\end{figure}

\begin{figure*}
\begin{center}
\includegraphics[width=14cm]{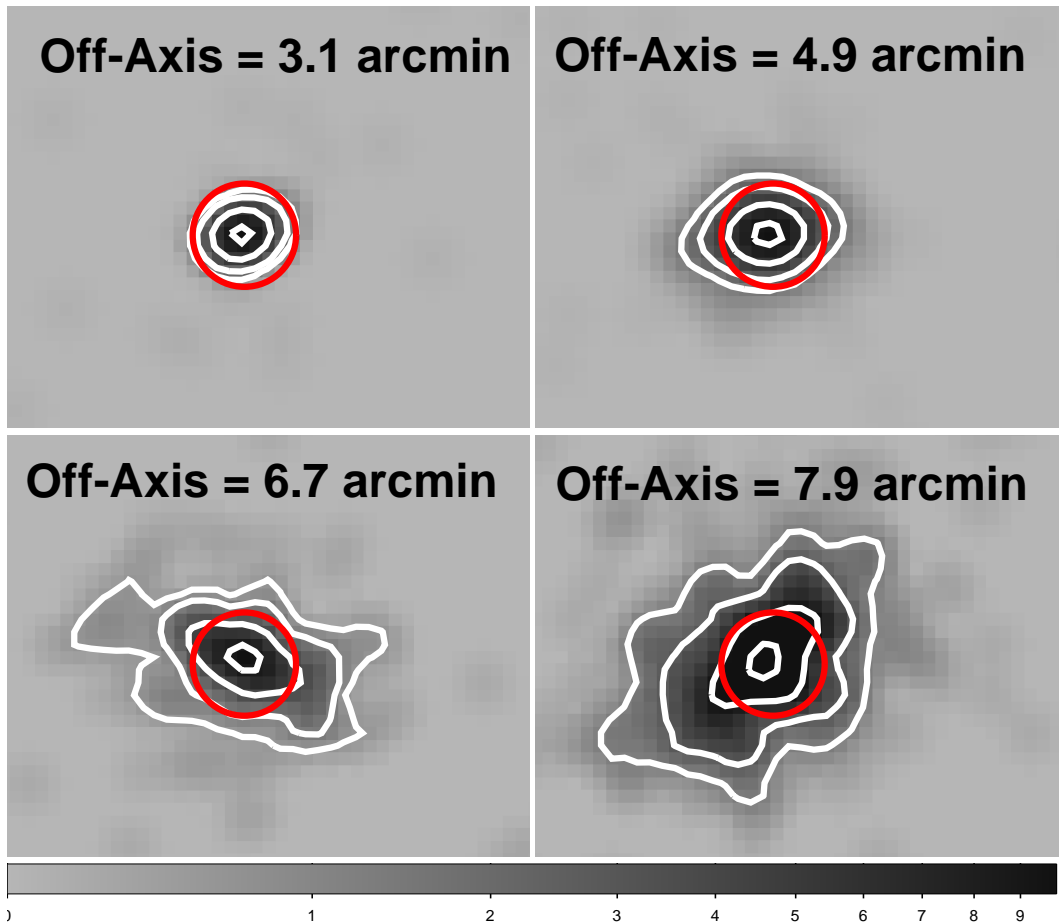}
\end{center}
\caption{The image of the same source (i.e., source-id 50 in the
C-COSMOS catalog presented in Paper I) observed in four overlapping
fields at different off-axis angles. The contours are drawn at 90\%,
50\%, 25\%, and 10\% of the peak counts. The red circles centered on
the position of the source have a radius of 2 arcsec.
\label{4psf}}
\end{figure*}

We give here a brief description of the observations and data
reduction. The full details are given in Paper I.  The C-COSMOS field
covers a contiguous area of $\sim$~0.92 deg$^2$, centered at 10$^h$
00$^m$ 18.91$^s$ +02$^\circ$ 10' 33.48'', near the center of the full
COSMOS field. The survey is made up of 36 different heavily overlapping
ACIS-I pointings, each with a mean exposure of $\sim$~50 ksec, for a
total exposure of 1.8 Msec. Twelve of the 36 pointings were scheduled
as two or more separate observations, with very similar roll-angles,
thus resulting in 49 observations in total.  Fig. \ref{overlay} shows the
number of ACIS-I pointings per pixel. Note that the central $\sim$~0.45
deg$^2$ area is covered by four to six overlapping pointings, while
most of the outer $\sim$~0.47 deg$^2$ area is covered by one to
two overlapping pointings. 
As an example, Fig. \ref{4psf} shows the
image of the same source observed in four overlapping fields at
different off-axis angles.

The 49 observations were processed using the standard CIAO 3.4
software tools\footnote{http://cxc.harvard.edu/ciao/} (Fruscione et
al. 2006).  Event files were cleaned of bad pixels, soft proton flares
and cosmic-ray afterglows, and were brought to a common reference
frame by matching the positions of bright X-ray sources with the
optical position of bright (18$<$I$<$23), point-like optical
counterparts. The systematic shifts between the X-ray and optical
positions are $\Delta$ RA$=$0.04'' and $\Delta$ DEC$=$0.25'' (see
Paper I). Observations with the same aim points and consistent
roll-angles were merged together, producing 36 event files, one for
each independent pointing.

\begin{figure}
\includegraphics[width=8.5cm]{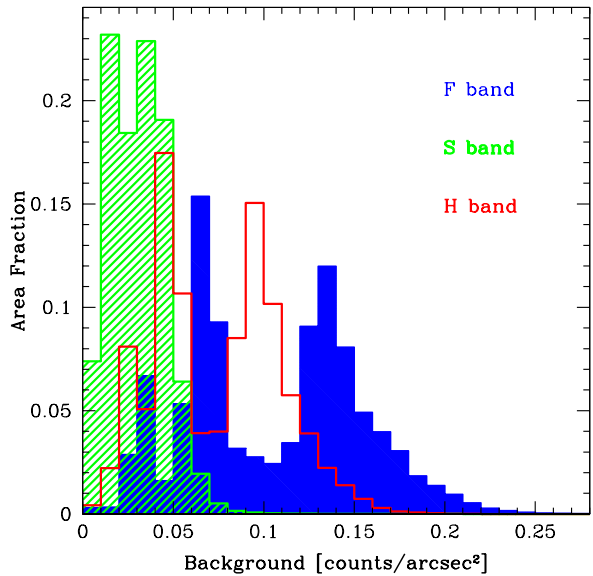}
\caption{ Area fraction for a given background per square arcsec in
the F band (solid blue histogram), S band (dashed green
histogram), and H band (empty red histogram).
\label{back}}
\end{figure}

The flux limits for source detection are influenced by three main
factors: (1) net exposure time, (2) background per pixel, and (3) size
of the source extraction region, which in turn depends on the size of
the PSF at the given position. The {\it Chandra} ACIS-I
on-axis PSF has a spatial resolution of 0.5''
FWHM, equivalent to $<$ 4-4.5 kpc at any redshift, and permits
observations of up to $\sim$~Msec to be photon limited. The adopted
tiling produces a rather homogeneous exposure time over the C-COSMOS
field (i.e., $\pm$12\% in the central $\sim$~0.45 deg$^2$ area) and a
uniform background. In the vignetting-corrected exposure time we
clearly distinguish two main peaks at 80 and 160 ksec (see Fig. 7 of
Paper I).  Fig. \ref{back} shows the fraction of the C-COSMOS area
with a given background per square arcsec in the three analyzed energy
bands: 0.5-7 keV (full band, F), 0.5-2 keV (soft band, S), and 2-7 keV
(hard band, H). We see two main peaks at 0.07 and 0.14
counts/arcsec$^2$ in the F band and at 0.02 and 0.04 in the S band,
corresponding to the two main peaks of the exposure time
distribution. These peaks correspond to a level of $\sim$~2 and
$\sim$~4 counts in the F band, and $\sim$~0.6 and $\sim$~1.2 counts in
the S band over an area of 3 arcsec radius, a typical source detection
region for off-axis angles less than 5-6 arcmin. Even the area with the
largest exposure time has therefore relatively low background for
point source detection; this is important for the detection of the
faintest sources.

\section{Generation of simulated data}

Extensive simulations were performed in order to test various source
detection schemes. The simulations were used (1) to test the
reliability of the source position reconstruction, (2) to verify the
count rate reconstruction, and (3) to assess and validate the level of
significance of each detected source at each given detection threshold
and thus to evaluate the level of completeness of the source list as a
function of flux.

\subsection{Creating the simulated input source catalog}

\begin{figure}
\includegraphics[width=8cm]{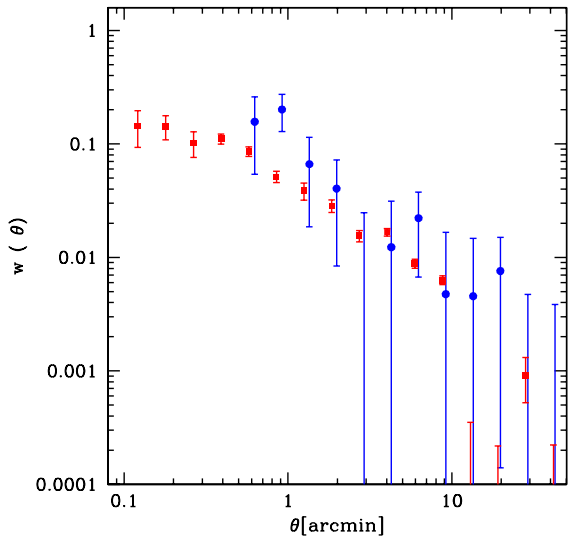}
\caption{The ACF of the sources selected from the COSMOS Mock catalog
for input to the C-COSMOS S band simulation (red squares)
compared with that of XMM-COSMOS (S band, blue solid dots).
\label{xmmacf_vs_mock}}
\end{figure}

In order to include realistic source clustering into the simulated
data, we sampled particles from a COSMOS Mock galaxy catalog (V3.0)
derived by Kitzbichler and White (2008). They made use of the
Millennium Simulation (Springel et al. 2005), a very large simulation
which follows the hierarchical growth of dark matter structures from
redshift z$=$127 to the present. The simulation assumes the
concordance $\Lambda$CDM cosmology and follows the trajectories of
2160$^3$ ($\sim$~10$^{10}$) particles in a periodic box 500 h$^{-1}$
Mpc on a side, using a special reduced-memory version of the GADGET-2
code (Springel et al. 2001; Springel 2005). The formation and
evolution of the galaxy population is simulated by using a
semi-analytical model (Croton et al. 2006, De Lucia \& Blaizot,
2007). We randomly selected 10000 mock galaxies per square degree in
the ad hoc redshift range 0.4$<z<$0.9 and i band magnitude range
17$<i<$26. The selected random sources in this redshift-magnitude
range show the same angular correlation function (ACF) as the S band
XMM-COSMOS sources (Miyaji et al. 2007), as shown in
Fig. \ref{xmmacf_vs_mock}, not taking into account that the
clustering strength could depend on the survey flux limit
(Plionis et al. 2008).
The agreement between the ACF of the random sample and the XMM-COSMOS
sample is good down to the 0.5 arcminute scale.  Below 0.5 arcmin, the
uncertainties in the S-band of the XMM-COSMOS ACF and the other X-ray
ACF from literature (see e.g., the {\it Chandra} Deep Field South,
D'Elia et al. 2004) are too large to allow them to be sensibly
compared with the one we derive. Each simulated galaxy was then
assigned an S band flux, randomly drawn from the number weighted $\log
N$ -- $\log S$ relation of the AGN population synthesis model by Gilli
et al. (2007). The corresponding minimum S band flux for the input
particles was $\sim~3\cdot 10^{-18}$ erg s$^{-1}$ cm$^{-2}$, which is
a factor 100 below the detection limit of C-COSMOS.  Hence background
fluctuations due to unresolved faint sources are included in the
simulations.

The S band flux of each source was then converted into an F band flux
assuming a power$-$law spectrum with an energy index
$\alpha_E =$0.4\footnote{f$_E$$\propto$E$^{-\Gamma}$, with $\Gamma = \alpha_E +1$}. The
simulated sources cover a 3 deg$^{2}$ sky area, which is enough to
completely enclose the COSMOS field.

\subsection{Creating the simulated X-ray event files}

Using the MARX simulator\footnote{http://space.mit.edu/CXC/MARX/} (version
4.2.1), we simulated a set of
49 {\it Chandra} ACIS-I pointings with the same exposure times, aim
points, and roll-angles as the real C-COSMOS pointings (see Paper
I). The simulated source list was fed into each simulated pointing and
net source counts F recorded. This procedure returns 49 {\it Chandra}
events files containing only source photons. 

To include a background appropriate to each pointing we used the CXC
compilation of blank sky
fields\footnote{http://cxc.harvard.edu/contrib/maxim/acisbg/}. These
blank fields lie at high Galactic latitude, away from soft bright
features such as the North Polar Spur, and have a median exposure of
$\sim70$ ks. Point-like and extended sources down to fluxes that would
be detectable in each exposure have been excluded, and the individual
exposures have been stacked into different blank sky files. We chose
the stacked blank sky file appropriate for ACIS-I data at the epoch of
our observations\footnote{http://cxc.harvard.edu/contrib/maxim/acisbg/data/\\acisi\_D\_01236\_bg\_evt\_010205.fits},
filtered to keep only photons detected in VFAINT mode
observations. This blank sky field has a total effective exposure of
$\sim 1.5$ Msec.

We then extracted 49 background event files by randomly resampling the
events out of the blank sky file scaling by the exposure time of each
observation.
Faint simulated sources with only a few counts would not be detected
and increase the background level by $\sim~5\%$ at
the depth of the blank sky observations. Since these faint, unresolved
sources are already included in the blank sky files, in order to avoid
counting them twice, we removed 5\% of the photons in each background
event file. 

The background files were then reprojected to the
coordinates of the real pointings by using the real aspect solution
files, and then combined with the corresponding source event
files. The final result is a set of 49 simulated ACIS-I fields that
closely mirror the actual 49 observations.

\section{Choosing the C-COSMOS source detection and characterization procedure}

In order to fully exploit the large and deep C-COSMOS coverage a
particular care had to be devoted to maximize areal coverage and
produce uniform depth; C-COSMOS used a complex tiling, with four
overlapping pointings in most of the central $\sim$~0.45 deg$^2$ area
with the best exposure, and two overlapping fields in most of the
surrounding area, covering additional $\sim$~0.47 deg$^2$ (see
Fig. \ref{overlay}). As a result, each source is observed at different
off-axis angles, $\theta_i$ (i.e., the distance of the source position
from the aim point in all overlapping fields), and thus with different
PSFs. For some sources in the central area the number of different
$\theta_i$ is as high as six. This mixture of PSFs requires addressing
three main issues: (1) maximizing the sensitivity when the PSF changes
so widely between different observations of the same source (from
$\sim$~1 arcsec to $\sim~10$ arcsec half power radius); (2) maximizing
the spatial resolution aimed to obtain the best source localization
and the effective deblendig; (3)obtaining accurate photometry, even in
cases of partly blended sources. To solve these issues a dedicate
analysis procedure was developed, and the simulations were used to
determine and validate it.

We tested sliding cell and wavelet algorithms to find and locate
source candidates, and both PSF fitting and aperture photometry. In
particular, we compared the results obtained using the SAS {\it
eboxdetect}\footnote{http://xmm.esac.esa.int/sas/8.0.0/eboxdetect/}
and {\it
emldetect}\footnote{http://xmm.esac.esa.int/sas/8.0.0/emldetect/}
tasks, used for the XMM-COSMOS survey (Cappelluti et al. 2009),
with those obtained using the {\it PWDetect} code (Damiani et
al. 1997) and CIAO {\it wavdetect}\footnote{http://asc.harvard.edu/ciao/ahelp/wavdetect.html} (Freeman et al. 2002).
We compared  {\it PWDetect} and CIAO {\it wavdetect}
on a data subset including 8 ACIS-I fields and found consistent results. We 
adopt the {\it PWDetect} as the main wavelet algorithm because of its much 
faster processing time (i.e., factor of 40$\div$50) with respect to CIAO {\it wavdetect}.

\subsection{ PWDetect}

The {\it PWDetect} code (Damiani et al. 1997) was originally
developed for the analysis of ROSAT data, and was then adapted for the
analysis of {\it Chandra} and XMM-Newton data. This method is
particularly well suited for cases in which the PSF is varying across
the image, as for {\it Chandra} images, since {\it PWDetect} is based
on the wavelet transform (WT) of the X-ray image, i.e., a convolution
of the image with a ``generating wavelet'' kernel, which depends on
position and length scale, that is a free parameter. For the {\it
Chandra} data, the length scale is varied from 0.35'' to 16'' in steps
of $\sqrt{}2$. This choice spans the range from the smallest to the
largest (for large $\theta_i$) {\it Chandra} PSFs.  Both radial and
azimuthal PSF variations are accounted for by {\it PWDetect}, which
first assumes a gaussian PSF and then corrects by a PSF shape factor,
calibrated on both radial and azimuthal coordinates.  {\it PWDetect}
was run on each of the 36 event files with a low significance level of
$\sim~10^{-3} $, to have entries with just 5 source counts (i.e., to
pick up most of the input sources). The catalogs of source candidates
from overlapping fields were then merged. The off-axis angle
$\theta_i$ is recorded and the source position measured at the
smallest $\theta_i$ (i.e., with the best PSF) is adopted as the
reported source position. If a candidate is not detected in one or
more of the overlapping fields, the count rate is computed at the
position of the source candidate and within a circle of radius R$_i$,
corresponding to 90$\%$ of the encircled count fraction of the PSF
($f_{psf}$\footnote{$f_{psf}$ indicates a fraction of the source
counts distributed in a circular area, following the PSF
shape.}$=$90$\%$) at $\theta_i$, as calibrated by the
CXC\footnote{http://cxc.harvard.edu/caldb/}. Finally, a mean count
rate, that is weighted by the count rate errors, is associated at each
source. Analysis of the simulated data showed that all candidates with
a wavelet size smaller than the PSF size and less than 5 counts are
spurious detections. These were then excluded from the candidate
catalogs.

\subsection{ EBOXDETECT and EMLDETECT}

Both {\it eboxdetect} and {\it emldetect} are part of the
XMM-Newton SAS package and are based on programs originally developed
for the detection in ROSAT images (see e.g., Voges et al. 1999). {\it
eboxdetect} is a standard sliding cell detection tool, which is run on
each of the 49 single observations. {\it eboxdetect} produces a list
of candidate sources down to a selected low significance level. The
list of source candidates is then passed to the {\it emldetect} task.
{\it emldetect} performs a simultaneous maximum likelihood PSF fitting
for each candidate to all the images at each position (see
e.g. Cappelluti et al. 2007 for more details on {\it eboxdetect} and
{\it emldetect}).{\it eboxdetect} was run setting a low significance level
(DET\_ML=3 or P$_{random}$=0.05), to provide a list of source candidates to {\it
emldetect}, that recognizes all possible significant sources.

{\it emldetect} has been adapted to run on {\it
Chandra} data by replacing the {\it XMM-Newton} PSF library with the
{\it Chandra} PSF library (see note 22), and to work with many
different PSFs, simultaneously. The counts at each position were
fitted using a model obtained by convolving the PSF at that position
with a $\beta$ model (Cruddace, Hasinger and Schmitt 1988). The
program interpolates over the calibration library of {\it Chandra}
PSFs to find the most appropriate PSF at the position of each source
in each observation. The more crowded is the field, the more
candidates are fitted simultaneously. {\it emldetect} can provide both
source positions and source count rates, or only source count rates
using fixed source positions. We ran it fitting for both source positions and count rates.
The best fit maximum
likelihood, DET\_ML, is related to the Poisson probability that a
source candidate is a random fluctuation of the background
(P$_{random}$):
\begin{equation}
DET\_ML=-ln(P_{random})
\end{equation}
Sources with low values of DET\_ML, and correspondingly high values of
P$_{random}$, are then likely to be background fluctuations.

\subsection{Tests on simulations}

We ran both detection algorithms on the simulated data. Catalogs of
candidates were produced with both {\it eboxdetect} and {\it
PWDetect}. These lists were visually inspected to identify obviously
spurious detections on the wings of the {\it Chandra} PSF around
bright sources, and near the edges of the ACIS-I chips. For both
detection algorithms, the number of these clearly spurious detections
is rather small in all three bands ($<1-2\%$). These entries were
deleted and the 'cleaned' lists used as input for the {\it emldetect} tool.
The {\it emldetect} output catalog was then cut at a conservative
value of DET\_ML=12 ($P_{random}<6\times10^{-6}$), to ensure that the
number of spurious detections in this catalog is practically zero, so that
the results are not contaminated by spurious associations.

Matched catalogs between the input simulated catalog and the {\it
emldetect} and {\it PWDetect} output catalogs were produced using two
methods: (1) a conservative approach, using a fixed matching radius of
0.5 arcsec. This produces matched catalogs which probably miss a
fraction of real associations, but are virtually free from spurious
associations. (2) A maximum likelihood algorithm, to find the most
probable association between an input source and an output detected
source. We used the catalogs produced using the first method to study
the accuracy of source localization and flux reconstruction, while we
used the catalogs produced by the second method to study the
completeness and reliability of the detection algorithms (see Sect. 5).

Table \ref{tabcomp} summarizes the comparison of the results of the
application of {\it eboxdetect}$+${\it emldetect} and {\it PWDetect} on simulated data.

\begin{deluxetable}{ccc}
\tabletypesize{\scriptsize} 
\tablecaption{Comparison between {\it eboxdetect}$+${\it emldetect} and {\it
PWDetect}}
\tablewidth{0pt} 
\tablehead{ \colhead{Parameter}  & \colhead{ {\it eboxdetect}$+$ {\it emldetect}}& \colhead{{\it
PWDetect} }  
     \\
\colhead{1 } & \colhead{ 2} & \colhead{3 } }
\startdata
\cutinhead{Comparison on source position:}
 $<$$\Delta$ R.A.$>$\tablenotemark{a}  & 0.17''$\pm$0.16''  & 0.02''$\pm$0.15''  \\
 $<$$\Delta$ Dec.$>$\tablenotemark{a}    & -0.18''$\pm$0.15 & 0.003''$\pm$0.15'' \\
 $\Delta$ R.A. RMS\tablenotemark{b}       &   0.32'' &  0.31'' \\
 $\Delta$ Dec. RMS\tablenotemark{b}     &  0.35'' & 0.34'' \\ 
\cutinhead{Comparison on completeness of close pairs:}
 $\%$ of missed pairs\tablenotemark{c}  & $\sim$~75\% &  \\
\cutinhead{Comparison on source photometry:}
 $<$ $F_x(F)\over{F_{xI}(F)}$ $>$\tablenotemark{d}  & 0.97$\pm$0.11  & 0.86$\pm$0.12  \\
$<$ $F_x(S)\over{F_{xI}(S)}$ $>$\tablenotemark{d}  & 1.00$\pm$0.12  & 0.94$\pm$0.14  \\
$<$ $F_x(H)\over{F_{xI}(H)}$ $>$\tablenotemark{d}  & 1.05$\pm$0.16  & 0.88$\pm$0.17  \\
 \enddata
 
\tablenotetext{a}{The median and interquartile of the shifts between the R.A. or Dec. of
the input sources and the R.A. or Dec of the detected sources, see also Fig. \ref{position}.}
\tablenotetext{b}{The RMS of the R.A. or Dec. shifts between input and detected positions, see also Fig. \ref{position}.} 
\tablenotetext{c}{Percentage of the pairs with a separation smaller
than 4 arcsec, that are missed in comparison to {\it PWDetect}, see also Fig. \ref{dd1pairs}.}
\tablenotetext{d}{The median and interquartile of the ratio between the output detected and input simulated count rates in the
F, S, and H band, see also Fig. \ref{comparerates}.}

\tablecomments{Column (1) shows the parameters used to test the
accuracy of source localization, the completeness un the recovery of
close pairs, and the flux reconstruction of the two detection
algorithm, which we used. Column (2) and (3) show the results for the
{\it eboxdetect}$+${\it emldetect} and the {\it PWDetect} algorithm,
respectively.}

\label{tabcomp}
\end{deluxetable}

\begin{figure*}
\begin{tabular}{cc}
\includegraphics[width=8cm]{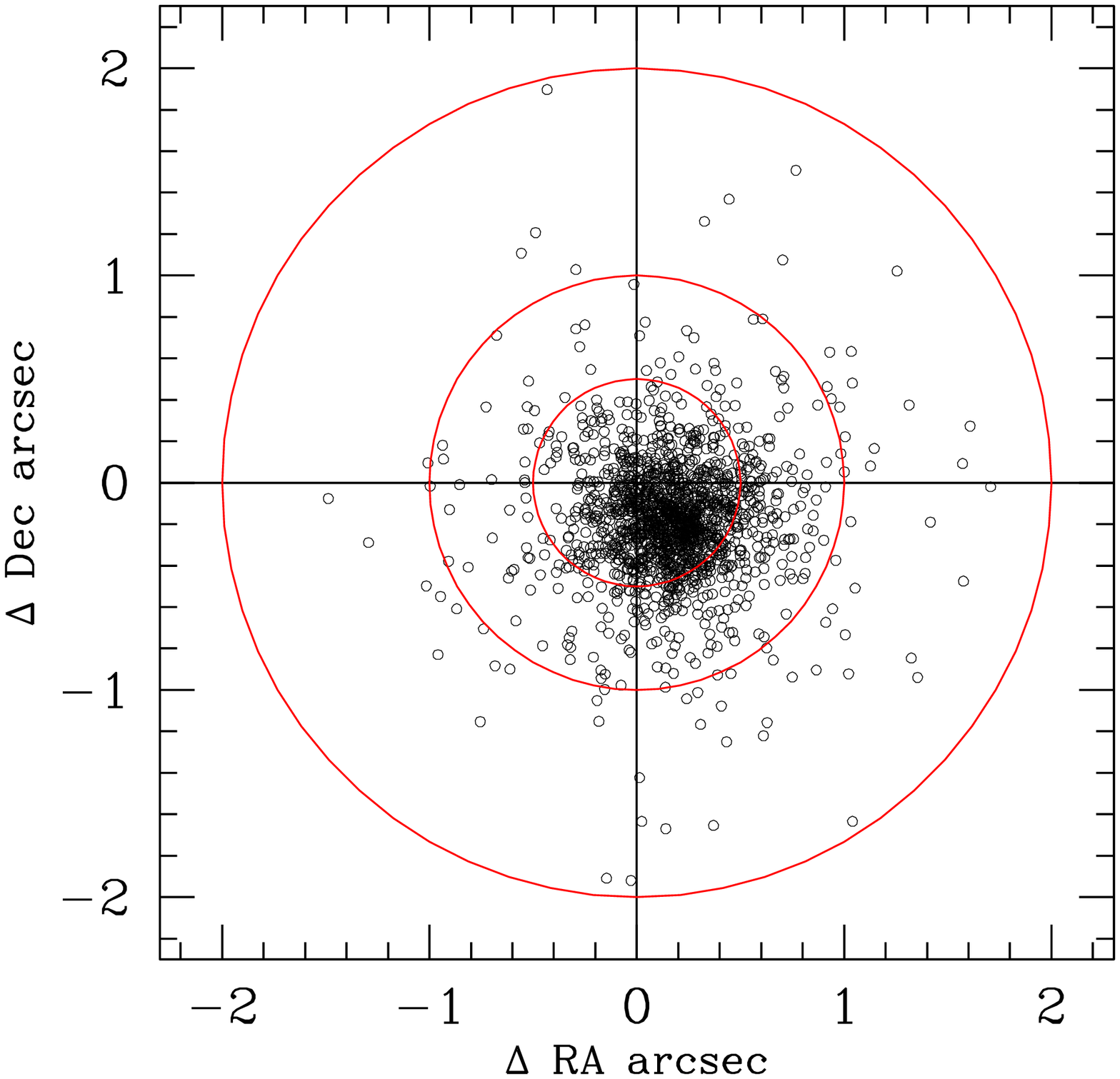}
\includegraphics[width=8cm]{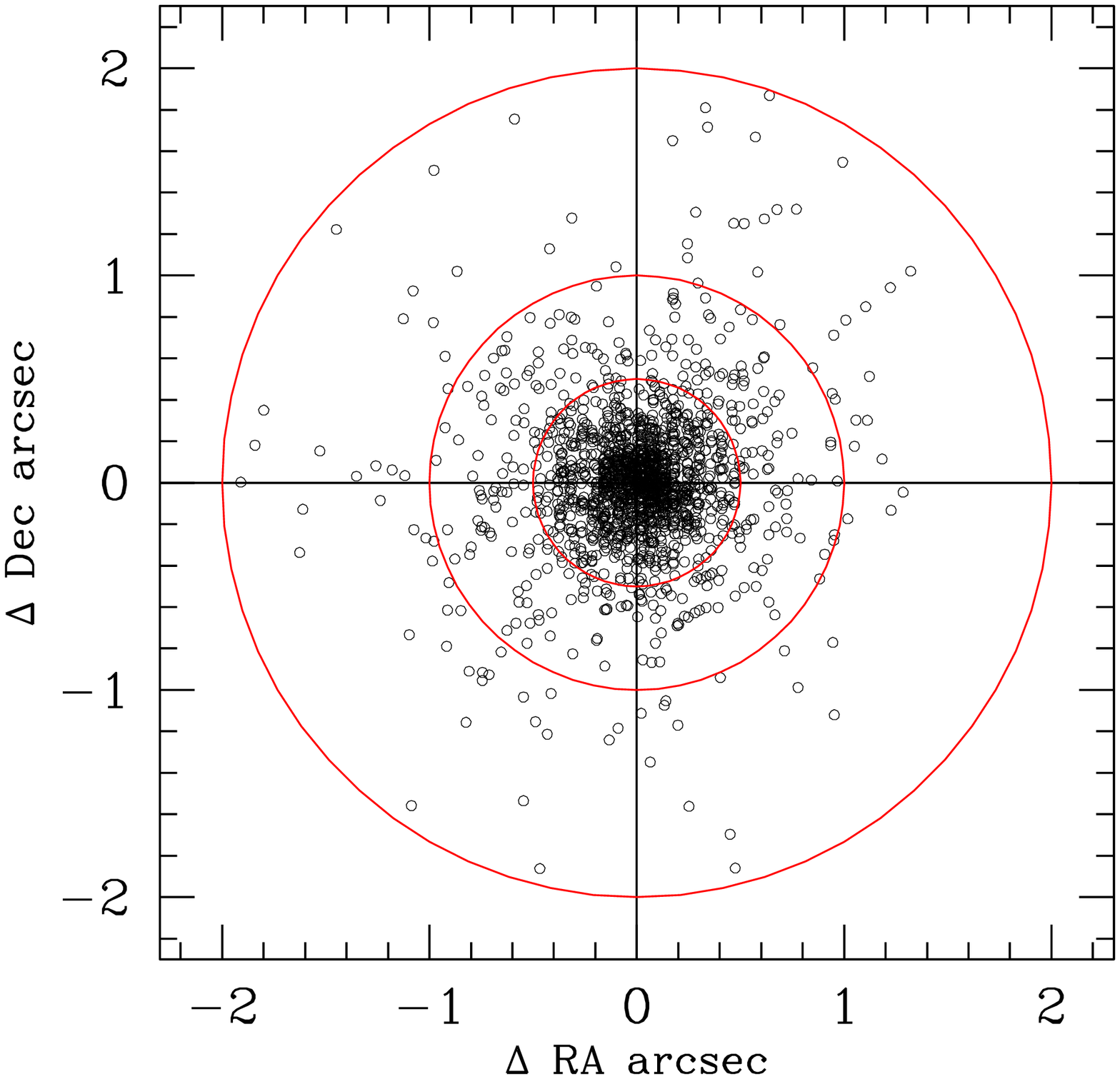}
\end{tabular}
\caption{ {\it Left panel:} shift between the input simulated source
positions and the source positions by {\it emldetect} using a matching
radius of 2 arcsec (black solid dots). The solid black lines represent
the zero shifts. The red circles have a radius of 0.5, 1, and 2
arcsec, respectively. {\it Right panel:} shift between the input
simulated source positions and the source positions by {\it PWDetect}
using a matching radius of 2 arcsec (black solid dots). Symbols as in
the left panel.
\label{position}}
\end{figure*}

We first compared the best-fit source coordinates provided by {\it
emldetect} and {\it PWDetect} with the input source positions (see
Tab. \ref{tabcomp}). The RMS variations and the interquartile of the
shifts are similar for the two detection algorithms; however, we
find a small systematic median shift between input and detected
R.A. and Dec. (see also Fig. \ref{position} ) using {\it
emldetect}. We conclude that {\it PWDetect} provides positions of
higher quality.

\begin{figure}
\includegraphics[width=8.5cm]{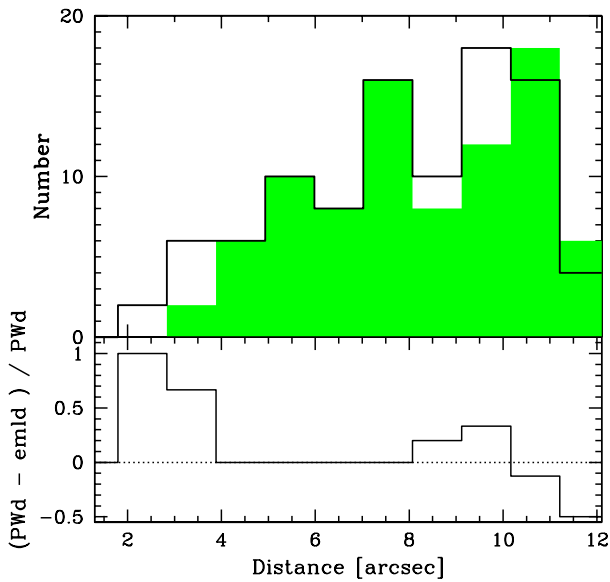}
\caption{{\it Top panel}: number of pairs in the F band detected by
{\it PWDetect} (black empty histogram) and {\it emldetect} (green
solid histogram), as a function of the separation.  {\it Bottom
panel}: ratio between the difference between the number of pairs
detected by {\it PWDetect} and {\it emldetect}, and the pairs detected
by {\it PWDetect} as a function of the separation.
\label{dd1pairs}}
\end{figure}

As a second step, we focussed on the ability of the detection
algorithms to separate close pairs of sources in {\it Chandra} data,
comparing the numbers of pairs found by {\it emldetect} and {\it
PWDetect} (see Fig. \ref{dd1pairs} and Tab. \ref{tabcomp}). The two
algorithms are equivalent for large ( $>$4'') separations, but there
is a deficiency in the number of pairs recovered by {\it emldetect} at
small ($<$4'') separations. We verified that all the $\sim$~75\% of
the pairs with a separation smaller that 4 arcsec missed by {\it
emldetect} are in the input source list, and not spuriously created by
the splitting of a single source. Analysis of the {\it eboxdetect}
candidate list and {\it emldetect} final list shows that the majority
($>$70\%) of these pairs are missed in the {\it emldetect} step, where
the program finds a best fit including one significant source only,
while the second falls below the detection threshold.  We conclude
that {\it PWDetect} is more efficient than {\it emldetect} at
resolving close pairs with separations $<$4'' and greater than
$\sim$~1.8''.

\begin{figure}
\includegraphics[height=6cm,width=9cm]{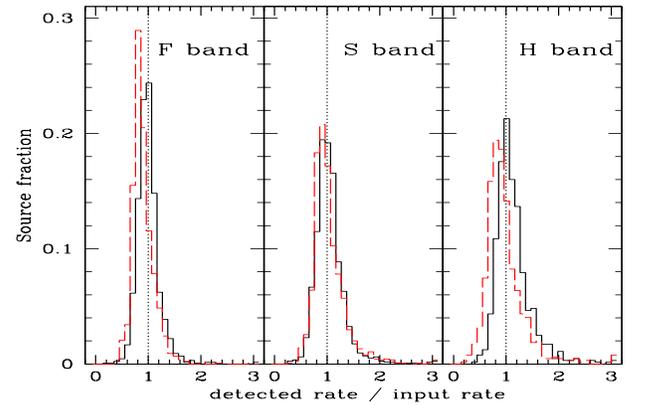}
\caption{The {\it PWDetect} (red dashed histogram) and {\it emldetect}
(solid black histogram) best fit count rates over the input count
rates in the F ({\it left panel}), S ({\it center panel}), and H (
{\it right panel}) band. The dotted vertical line corresponds to the exact match
between the evaluated count rates and the input count rates. Note as
the {\it emldetect} count rates are in good agreement with the input
count rates.
\label{comparerates}}
\end{figure}

Finally we compared the {\it emldetect} and {\it PWDetect} best-fit
count rates with the input count rates in the F, S, and H band (see
Fig. \ref{comparerates} and Tab. \ref{tabcomp}). The {\it PWDetect}
reconstructed count rates were systematically smaller than the input
count rates by 10-20\%.  A similar problem was found by Puccetti et
al. (2006) using a similar detection algorithm on {\it XMM-Newton}
data. {\it emldetect} reconstructs much better the count rates in all
the bands.

\begin{figure*}
\begin{tabular}{cc}
\includegraphics[width=8cm]{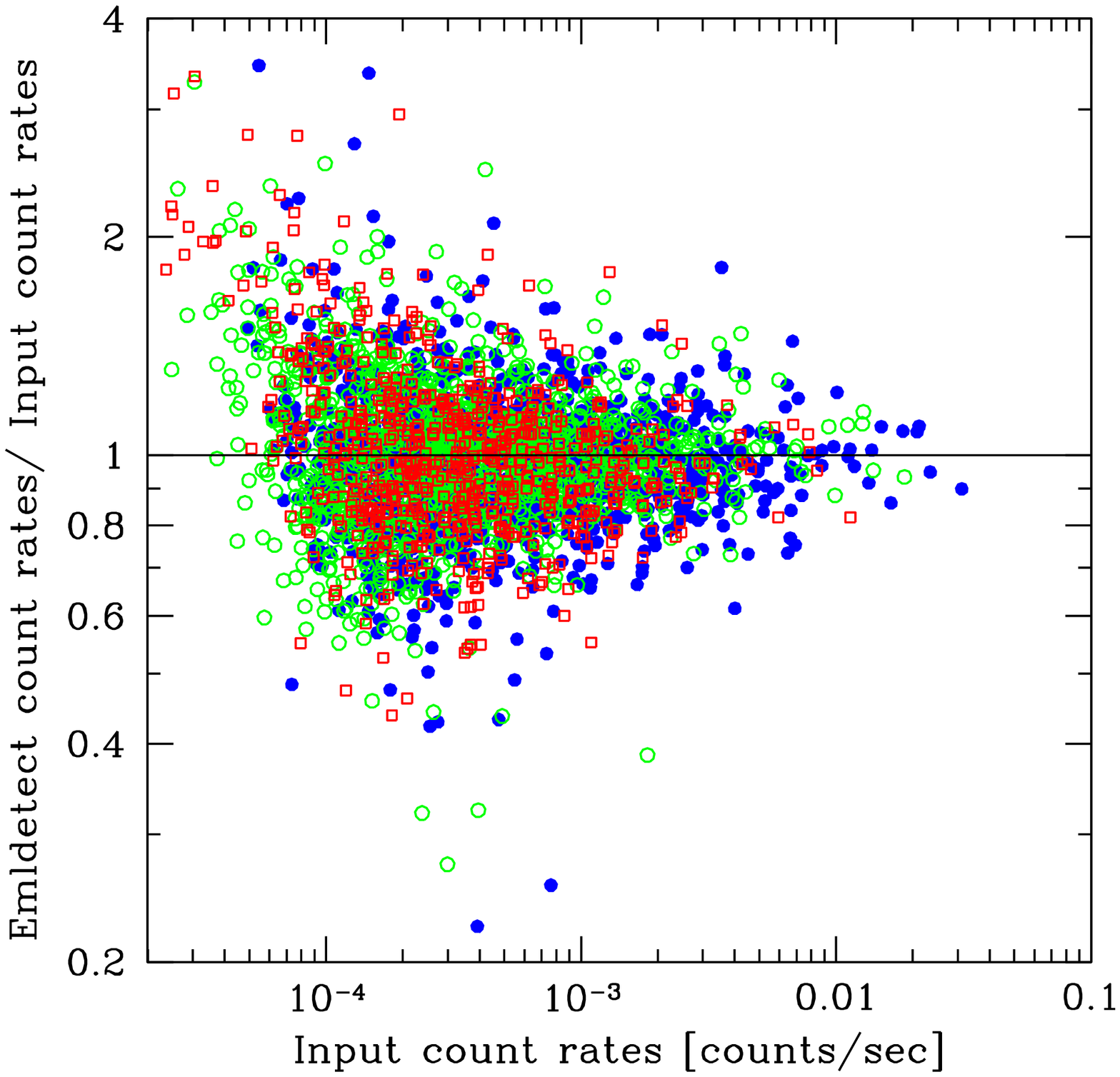}
\includegraphics[width=8cm]{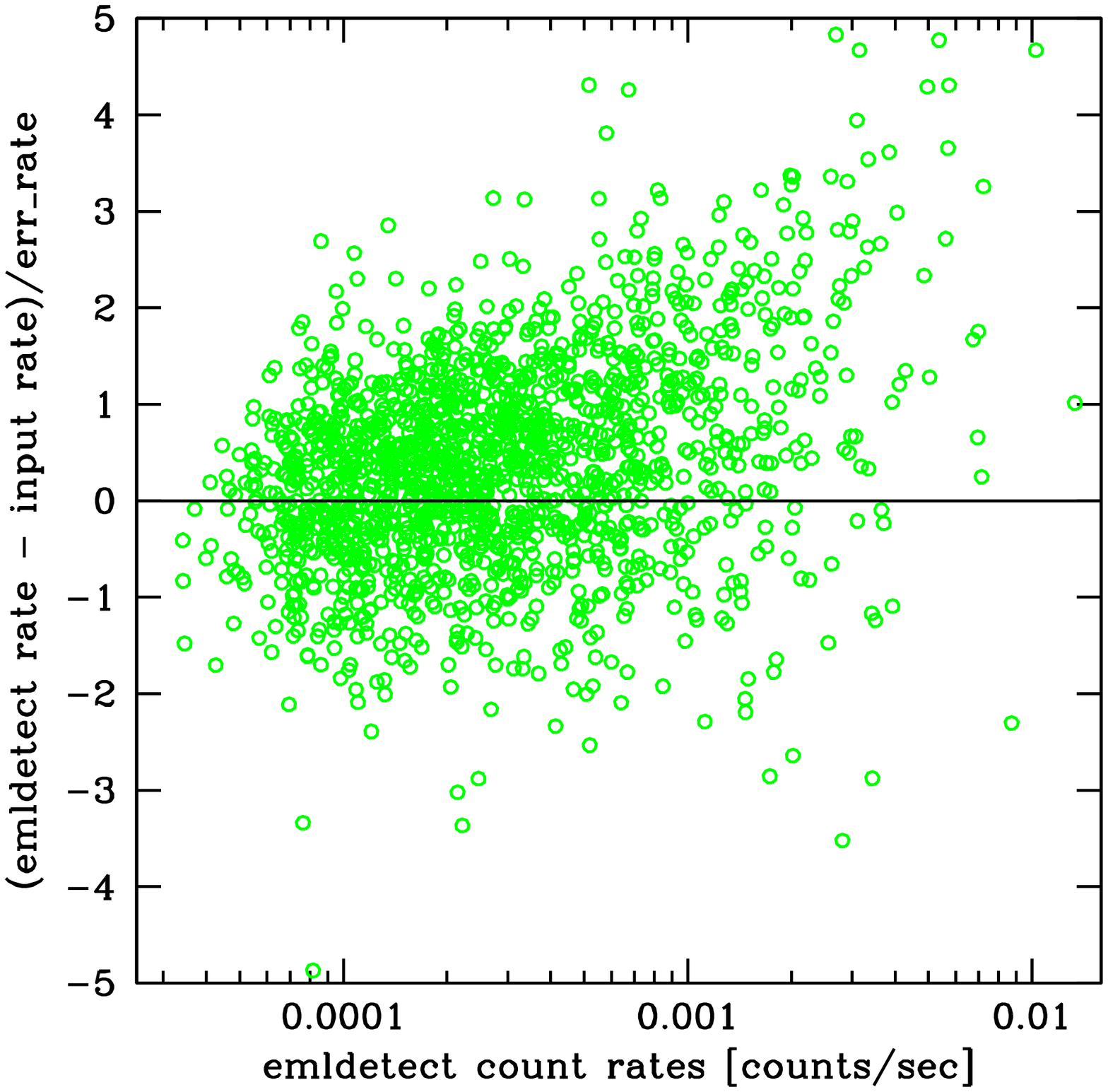}
\end{tabular}
\caption{{\it Left panel:} The ratio between the best fit count rates obtained by {\it
emldetect} and the input count rates as a function of
the input count rates for the simulations in the F (blue solid
circles), S (green open circles), and H (red open squares) band. The
solid line is the exact match between the best fit count rates and the
input count rates. {\it Right panel:} the difference between the {\it
emldetect} count rate and the input count rate divided by the {\it
emldetect} error on the count rate, as a function of the count rate
for the sources detected in the S band.
\label{cratesim}}
\end{figure*}

The accuracy of the count rate reconstruction of the {\it emldetect}
algorithm is also good at all count rates, without any large
systematic shifts, both at low count rates and at high count rates
(see left panel of Fig. \ref{cratesim}). The right panel of
Fig. \ref{cratesim} shows the difference between the {\it emldetect}
count rate and the input simulated count rate divided by the {\it
emldetect} error on the count rate as a function of the {\it
emldetect} count rate. We see that the distribution is approximately
centered around zero for count rates smaller than $\sim$~0.5
counts/ksec, but becomes positive for larger count rates. This
suggests that at high count rates, there is a not negligible
systematic error in the {\it emldetect} count rate determination, due
to the uncertainties in the PSF model becoming comparable to, or
higher than, the statistical error. For this reason we also performed
aperture photometry (see Sect. 6.4), which should be free from this
systematic error.

\subsubsection{Error on the positions}

The source positional error is proportional to the PSF at the position
of the source, and inversely proportional to the square root of the
source number counts. We evaluated the errors on the positions by
dividing the PSF$_{radius}$ by (1) the square root of the total source
plus background counts (T, $Pos_{Error}= PSF_{radius}/\sqrt T$) and
(2) the square root of the net, background subtracted, source counts
(C$_s$, $Pos_{Error}= PSF_{radius}/\sqrt C_s$). We used different
PSF$_{radius}$, from 50\% to 90\% of the f$_{psf}$ at the position of
each source in the field where the source is detected at the smallest
$\theta_i$ (i.e., with the best PSF).

\begin{figure}
\includegraphics[width=8.5cm]{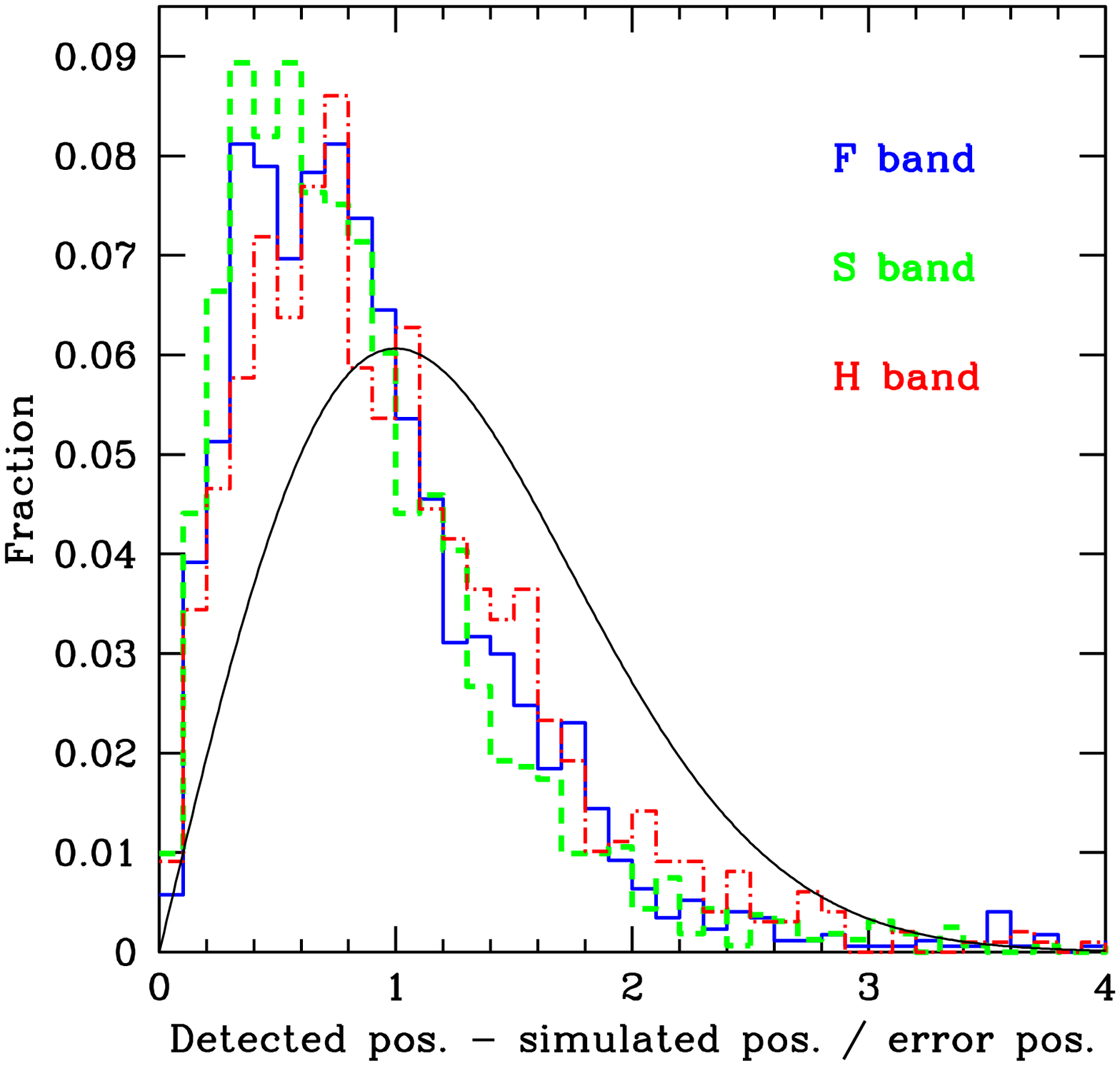}
\caption{The distributions of the ratio between the deviation between
detected positions by {\it PWDetect} and input positions and the X-ray error on the
position for the simulations in three energy bands (blue: F band;
green: S band; red: H band).  The solid curve is the expectation
based on Gaussian statistics.
\label{possim}}
\end{figure}

These errors were then compared with the deviations between the X-ray
positions and input positions in the simulations.  Method (2) gave the
best match using a PSF$_{radius}$ corresponding to the 50\% f$_{psf}$
at the $\theta_i$ of each source, and the counts included in a
circular region with the same radius. We used the f$_{psf}$ -
$\theta_i$ calibration provided by CXC (see note 19). Larger
PSF$_{radii}$ provided implausibly large position errors for bright
sources. Including background counts (method 1) produces too small
errors for faint sources, where the background is not negligible. For
$\sim$~60 sources with more than $\sim$~120 counts, the errors on
RA and Dec are formally smaller than 0.07 arcsec (i.e., errors on
source position smaller than 0.1 arcsec). In these cases the error on
source position was conservatively set to 0.1 arcsec to account for
possible small systematic errors in the astrometric corrections (see
Sect. 2 and 4.3).  Fig. \ref{possim} shows the distribution of the
ratio between the deviation between the {\it PWDetect} positions and
input positions and the X-ray error on the position evaluated as in
method (2). The distributions in the three detection bands are similar
and peak at a value of $\sim$~0.7-0.8. These distributions are
compared with the expectation based on Gaussian statistics which peaks
at unity. This comparison shows that the assumed errors on the
positions, although very small, are, on average, somewhat larger than
the deviation between input positions and detected positions. However,
to account for small systematic errors in the astrometric corrections,
which are not included in the input positions while they certainly
affect the observed data, we use in the following the conservative
errors on the positions computed as described above.

\subsection {The final C-COSMOS source detection and characterization
procedure}

In summary, the comparison of the two methods {\it PWDetect} and {\it
eboxdetect}$+${\it emldetect} on the simulated C-COSMOS field shows
that {\it PWDetect} is superior in separating closely spaced sources
and in localizing sources, and relatively poor at
photometry. Conversely, {\it emldetect} is poor at separating closely
spaced sources, while it is good at estimating source
reliability, completeness, and photometry.  These results suggested the
following source detection and characterization procedure:

\begin{itemize}

\item[1-] {\it PWDetect} is run first with a low threshold to produce a catalog of source
candidates, with the best localization;

\item[2-] this catalog of source candidates is used as input for {\it emldetect} which
performs a PSF fitting to find the best fit maximum likelihood source
count rate and the probability that each source candidate is a
fluctuation of the background. In {\it emldetect} the coordinates used to 
fit each source are those provided by {\it PWDetect} for the most on-axis observation;

\item[3-] aperture photometry is used to get good photometry for bright sources.

\end{itemize}

This combined approach allows us to obtain both the best possible
position determination and reliable photometry for all sources.

\section{Completeness and reliability}

\begin{figure*}
\begin{tabular}{ccc}
\includegraphics[width=6.2cm]{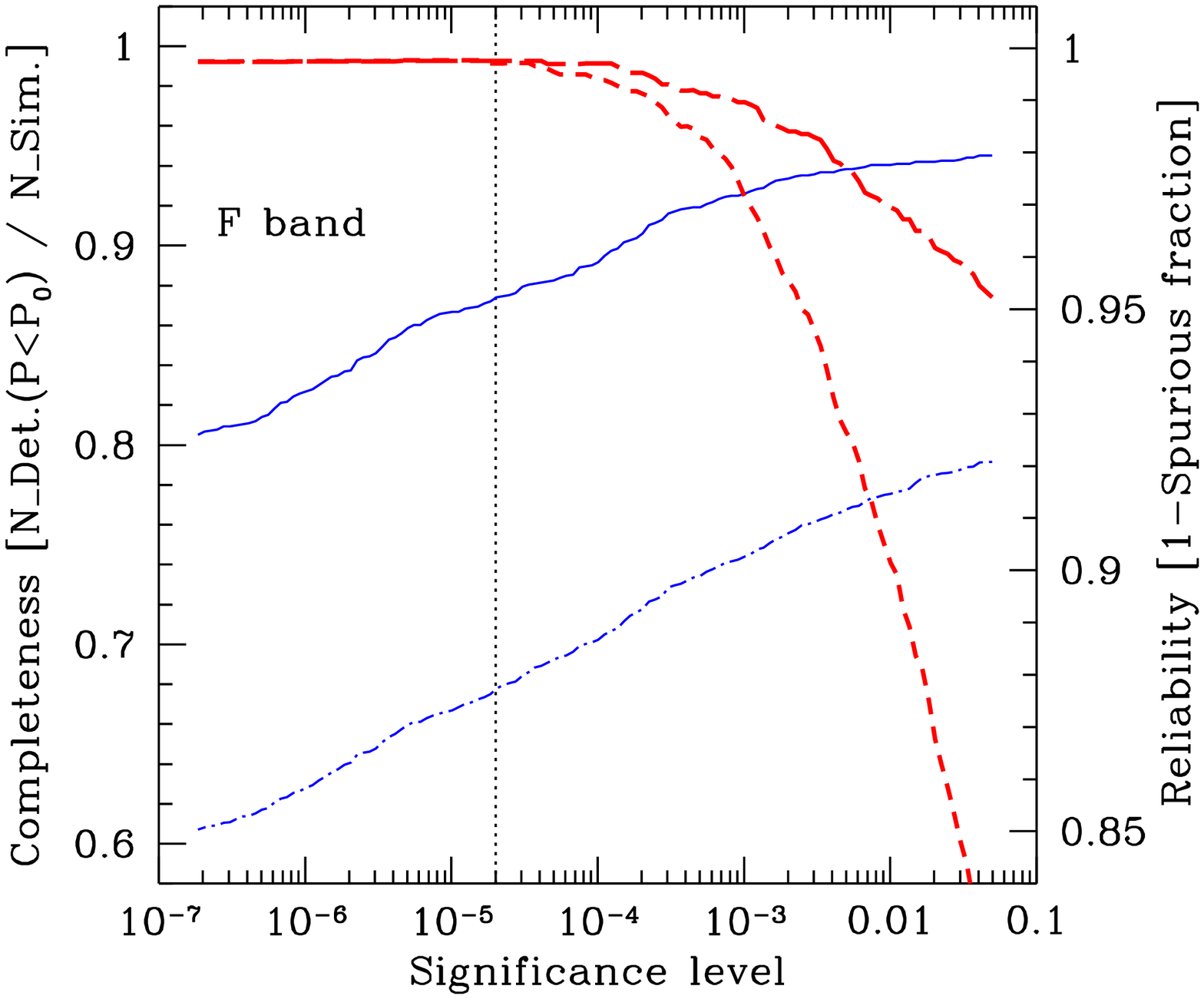}
\includegraphics[width=6.2cm]{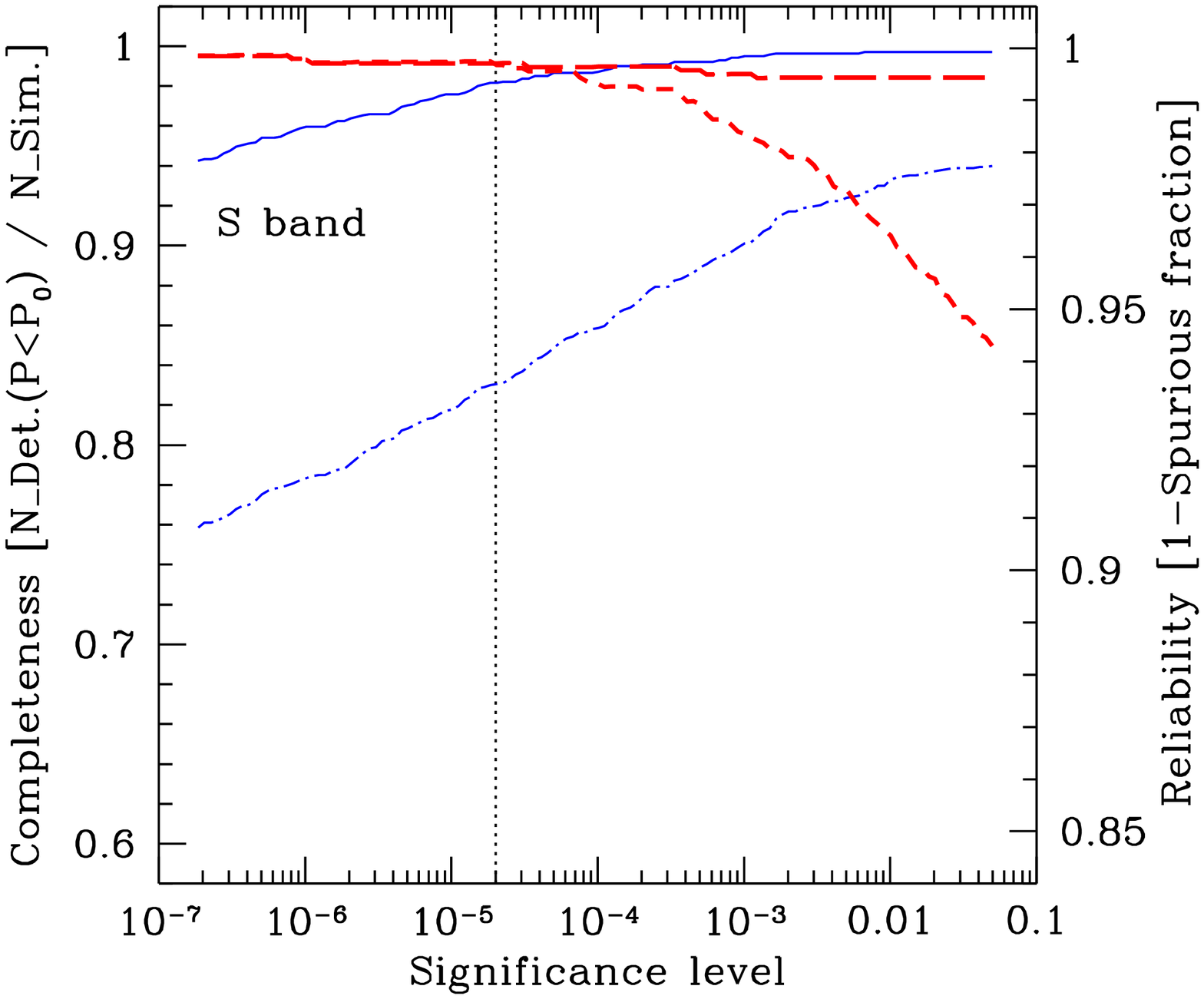}
\includegraphics[width=6.2cm]{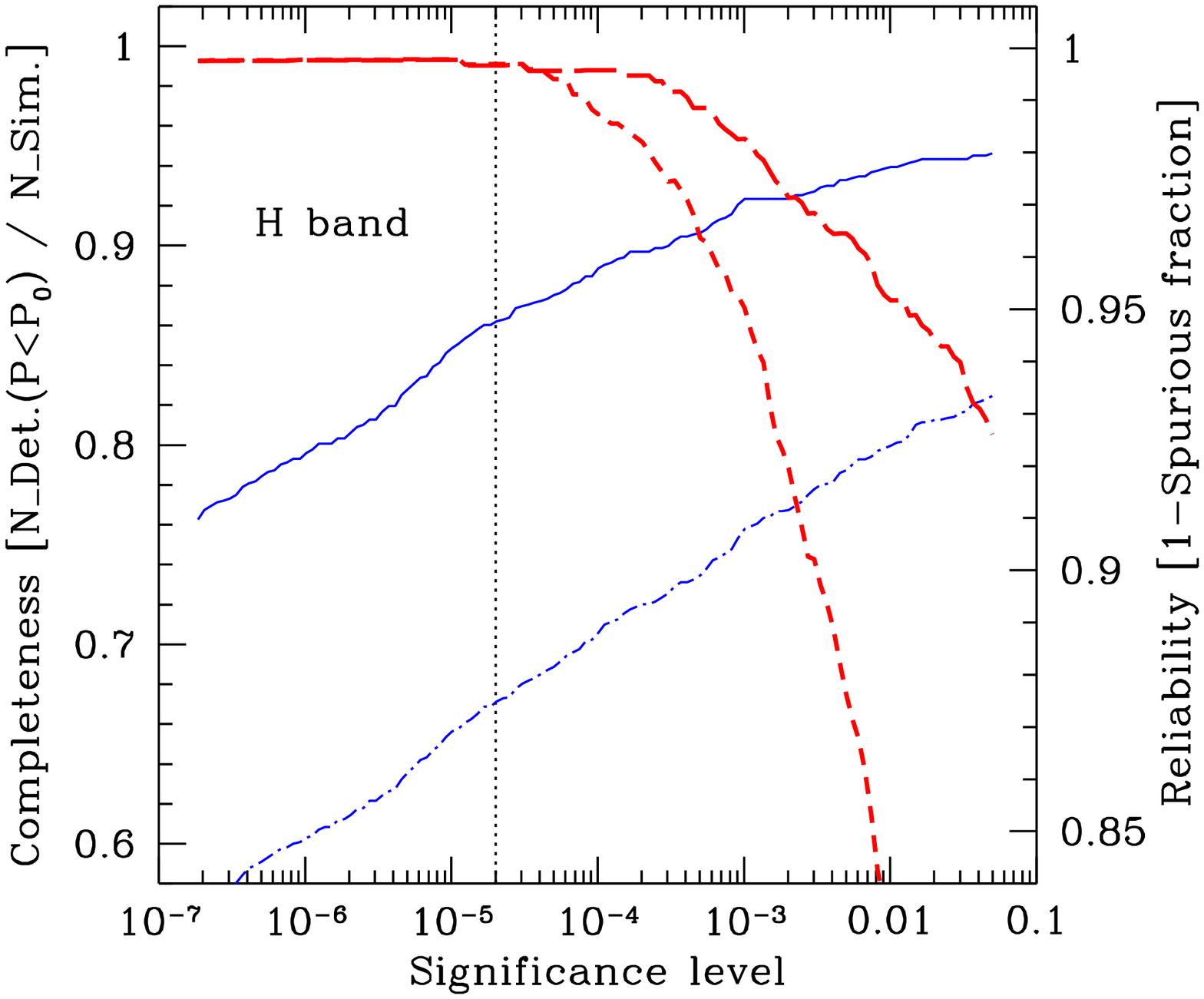}
\end{tabular}
\caption{Completeness (solid and dot-short dashed lines, left y axis) and reliability
(long dashed and short dashed lines, right y axis) as a function of the significance level for the simulations in the F ({\it left panel}), S ({\it center panel})
and H ({\it right panel}) band, for sources with at least 12 counts (solid and long dashed lines, respectively) and
at least 7 counts (dot-short dashed and short dashed lines, respectively). The dotted vertical black lines indicate the
chosen significance level of $2\cdot 10^{-5}$.
\label{complet} }
\end{figure*}

The threshold for source detection must be set by balancing {\em
completeness} (the fraction of true sources detected, i.e., ratio
between the number of the detected sources and the number of input
simulated sources) versus {\em reliability} (one minus the fraction of
spurious sources detected, i.e., one minus the ratio between the
number of spurious sources and the number of input simulated
sources). Our simulations allow us to choose a threshold which has a
known completeness and reliability. The three panels of
Fig. \ref{complet} show the completeness in the F, S, and H band as a
function of the significance level for sources with at least 12 counts
(solid lines) and 7 counts (dashed lines). The latter value refers to
the counts of a typical source close to our flux limit, where we
expect a rather large incompleteness. The former value (12 counts)
ensures significantly higher completeness. Fig. \ref{complet} also
shows the reliability as a function of the significance levels for the
same two cases. We chose a significance level of $2\cdot 10^{-5}$ (or
DET\_ML=10.8), which represents a reasonable compromise between high
completeness and high reliability. Higher significance levels give
higher completeness but lower reliability. At the chosen threshold we
have 87.5\% and 68\% (F band), 98.2\% and 83\% (S band), 86\% and 67\%
(H band) completeness for sources with at least 12 and 7 counts,
respectively. At the same significance level and the same counts
limits, the reliability is
$\sim$~99.7\% for the three bands and both source count limits. This implies about 5, 4, and 3 spurious
detections with $\geq7$ counts in the F, S, and H bands, respectively,
and 3, 4, and 3 spurious detections with $\geq12$ counts in the same
bands.

\begin{figure}
\includegraphics[width=8.5cm]{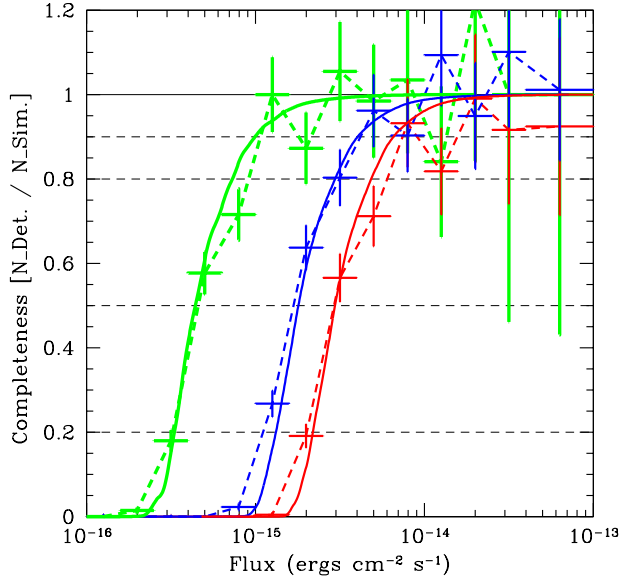}
\caption{The crosses represent the completeness as a function of
the flux at the chosen significance level of $2\cdot 10^{-5}$, in
F band (blue crosses), S band (green crosses), and H band (red
crosses). The dashed lines connect the relative cross points. The
solid lines represent the sky-coverage calculated as in Sect. 7.2 and
normalized to the maximum sky-coverage. The horizontal black dashed
and solid lines indicate 5 completeness fractions.
\label{complet2}}
\end{figure}

Fig. \ref{complet2} shows the completeness for a significance level of
$2\cdot 10^{-5}$ as a function of the flux for the F, S, and H
bands. Table 2 gives the flux limits corresponding to 4 completeness
fractions in the F, S, and H bands.

\begin{figure}
\includegraphics[width=8.5cm]{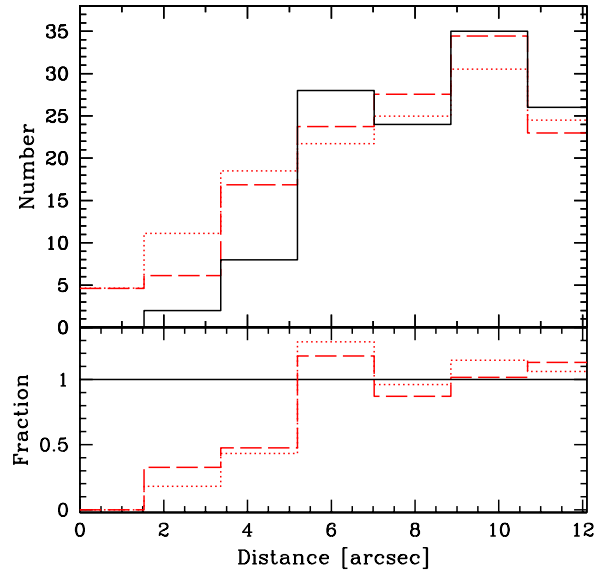}
\caption{{\it Top panel}: the number of pairs in the F band detected
by {\it PWDetect} (black solid histogram), compared to the number of
pairs in the simulations having one member with at least 7 counts (dotted histogram) and
12 counts (dashed histogram) as a function of the separation.  {\it
Bottom panel}: ratio between the pairs detected by {\it PWDetect} and
the number of pairs in the simulations with 7 or 12 counts as a
function of the separation. In both panels the numbers of pairs in the
simulations are corrected dividing them by the square of
the completeness expected at their counts thresholds (see
text).
\label{dd1pairs2}}
\end{figure}

We have also evaluated the completeness of the method in the detection
of close pairs. Fig. \ref{dd1pairs2} compares the number of pairs having one member with
at least 7 and 12 counts in the simulated data with the detected
number of pairs. The number of pairs in the simulated data have been
corrected dividing them by the square of the completeness expected at their
counts thresholds (87.5\% for the pairs with at least 12 counts and
68\% for the pairs with at least 7 counts). In fact to correctly
compare the number of pairs in the simulated data and the detected
number of pairs, it is necessary to take into account that the
detected number of pairs is not complete at the chosen significance
level, and moreover that each pair must be corrected for the
completeness of both sources in pair, that is the square of
completeness. We see that at distances smaller than 5 arcsec, we miss
between 50\% and 70\% of the pairs with at least 12 counts and between
70\% and 80\% of the pairs with more than 7 counts. The reason is that
it is increasingly difficult to detect a faint (7 or 12 counts) source
near a bright source, because of the wings of the PSF of the
latter. Indeed, all pairs recovered have a counts ratio $<3$, while
about 40\% of the input pairs have a count ratio $>3$, none of which
are detected in our analysis.

\begin{deluxetable}{lccc}
\tabletypesize{\scriptsize} 
\tablecaption{Flux limit and Completeness}
\tablewidth{0pt} 
\tablehead{ \colhead{Completeness } & \colhead{F(0.5-10 keV)  } & \colhead{ F(0.5-2 keV) } & \colhead{F(2-10 keV) } \\
    \colhead{\% } & \colhead{ \cgs} & \colhead{\cgs} & \colhead{ \cgs}
   } 
\startdata
90 & $4.1\cdot10^{-15}$ & $1.1\cdot10^{-15}$ & $7.8\cdot10^{-15}$ \\
80 & $3.1\cdot10^{-15}$ & $9.4\cdot10^{-16}$ & $6.1\cdot10^{-15}$ \\
50 & $1.7\cdot10^{-15}$ & $4.5\cdot10^{-16}$ & $2.9\cdot10^{-15}$ \\
20 & $1.1\cdot10^{-15}$ & $3.3\cdot10^{-16}$ & $2.0\cdot10^{-15}$ \\ 
 \enddata 
\label{tab1}
\end{deluxetable}

\section {Observed data: source detection and count rates}

Source detection and characterization were performed on the real,
observed, event files using the approach described in Sect. 4.4. The
three energy bands, F, S, and H were used. The candidate catalogs
produced by {\it PWDetect}, used as input for {\it emldetect} were cut
at a low threshold of $\sim~10^{-3} $, corresponding to 5 counts. The
number of {\it PWDetect} source candidates in each of the three bands
was between 2500 and 3500. These lists were visually cleaned to
identify obviously spurious detections on the wings of the {\it Chandra}
PSF around bright sources and near the edges of the ACIS-I chips,
following the same procedure adopted for the simulated data (see
Sect. 4). As for the simulations, the number of clearly spurious
detections is small in all three bands ($<1-2\%$).  

At the chosen probability (i.e. significance level $2\cdot10^{-5}$ or
DET\_ML=10.8), the number of spurious detections is presumably $<<$ 12
in the total catalog (i.e. F, S, and H band). The total catalog is
obtained by the cross-correlation of the three single band (i.e. F, S,
and H) catalogs, in this way the number of spurious sources in the
single F, S, and H bands, evaluated by the detailed analysis of the
simulations (see Sect. 5), are no longer indipendent. As a result the
number of the total spurious sources is less than the sum of the
spurious sources in each of three single bands.

\subsection{Source position}

\begin{figure*}
\begin{tabular}{cc}
\includegraphics[width=8.5cm]{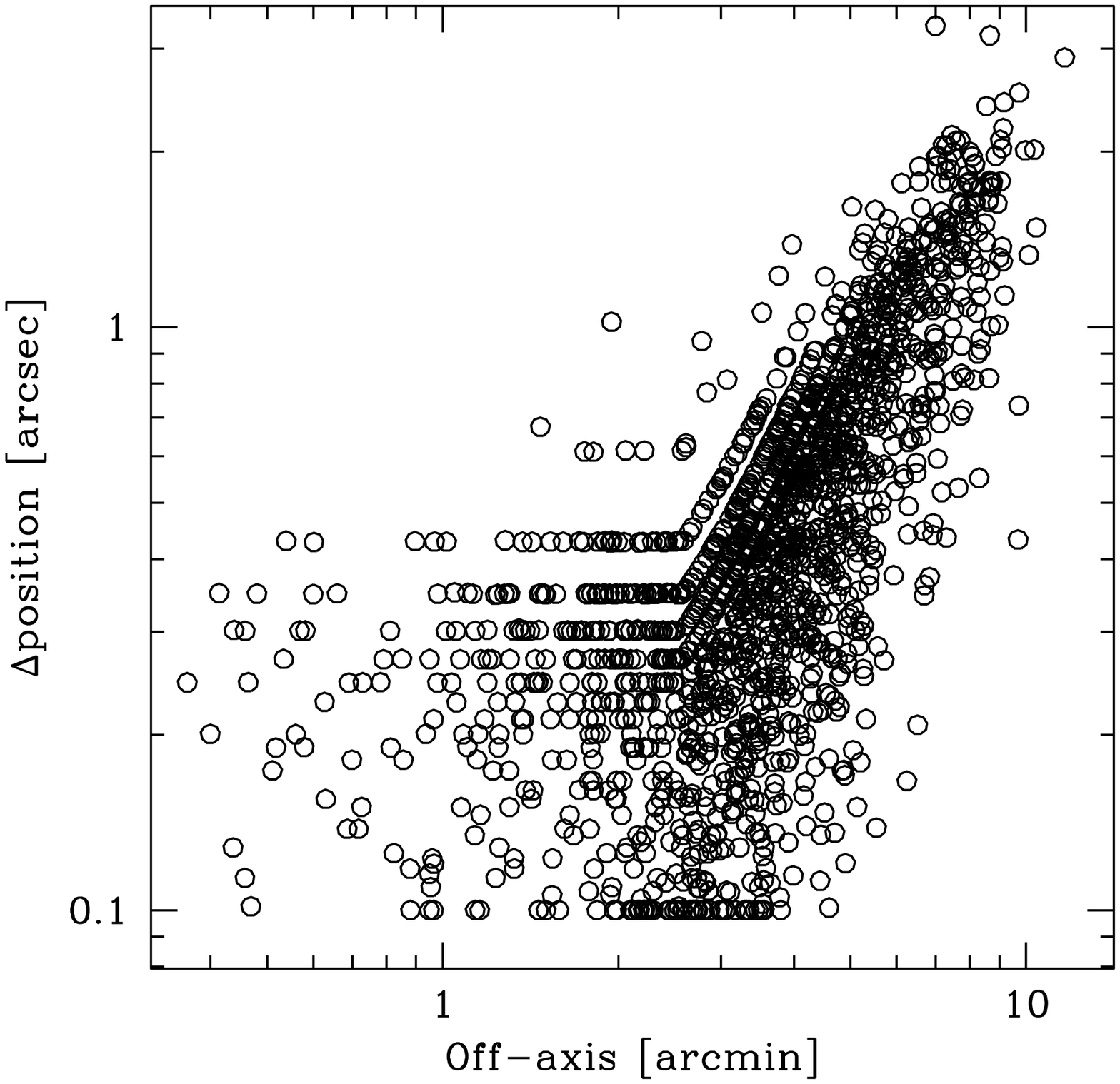}
\includegraphics[width=8.5cm]{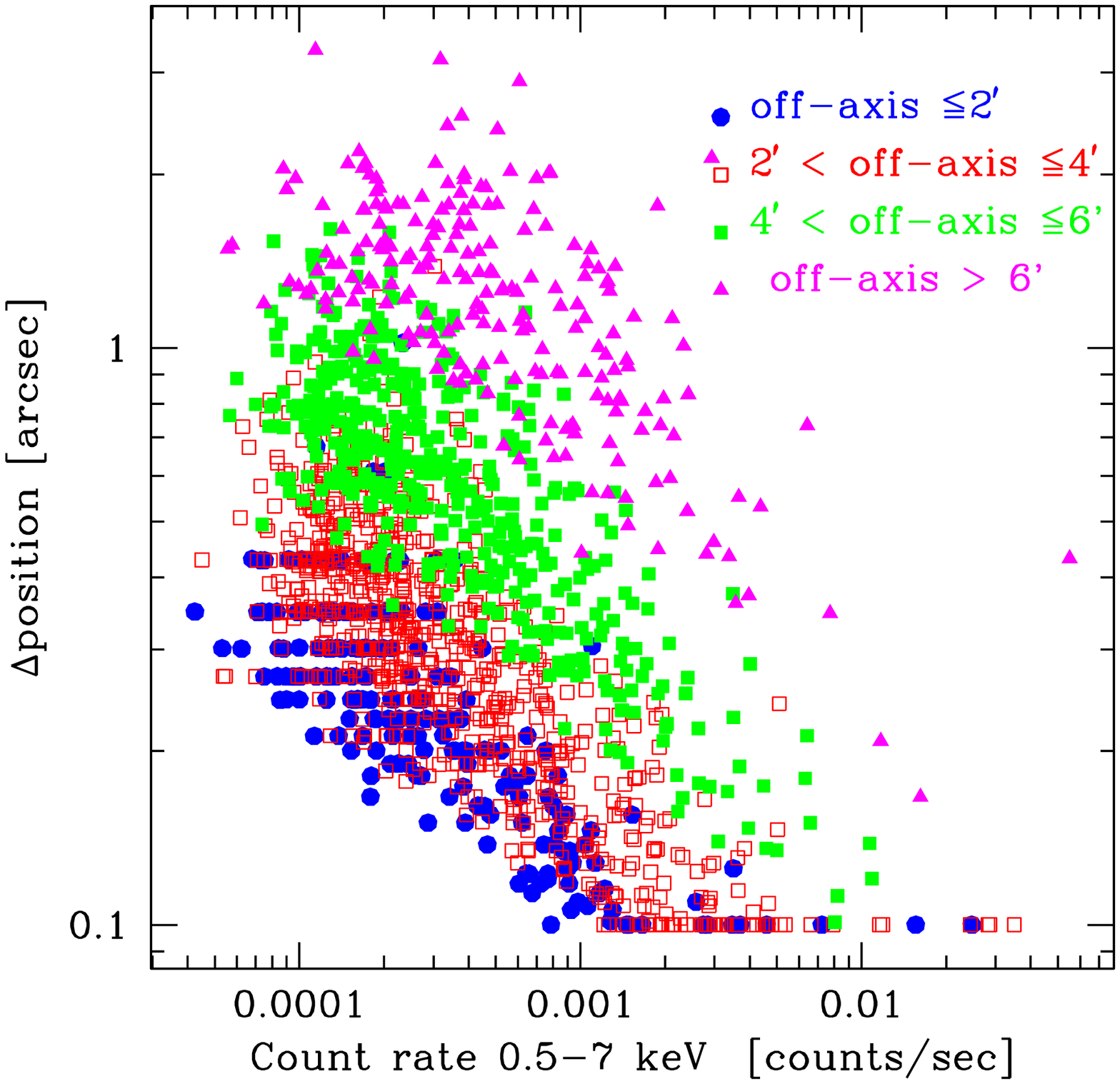}
\end{tabular}
\caption{{\it Left panel:} The error on the source position as a
function of the off-axis angle for sources detected in the F band.
{\it Right panel:} the positional error in the F band as a function of
the source count rate in 4 off-axis bins: filled circles = off-axis
$<$2$^\prime$; open squares = 2$^\prime<$off-axis$<4\prime$; filled
squares = 4$^\prime<$off-axis$<$6$^\prime$; filled triangles = off-axis
$>$6$^\prime$. The off-axis angle is the distance of the sources
candidate position from the aim point of the pointing where the
source position is measured with the best PSF (see Sect. 4.1).
\label{poserr}}
\end{figure*}

 Fig. \ref{poserr} (left panel) shows the positional error, evaluated
using the empirical technique described in Sect. 4.3.1, as a function
of the off-axis angle. The notch in figure depends on the fact that at
a fixed off-axis angle, the PSF$_{radius}$ is $\sim$constant, while
$\sqrt{C_s}$ is a discrete variable, since T are integer numbers and B
are small. The error is typically less than $\sim$~0.5 arcsec at the
smallest off-axis angles, $\theta_i$ $<$2 arcmin, and then increases
to 1-2 arcsec for $\theta_i$ $\ge$2 arcmin. Most of the scatter at a
given off-axis angle in this figure is due to the range of count rates
in the sources. Fig. \ref{poserr} (right panel) shows the positional
error in the F band as a function of the source count rate in 4
off-axis bins. Both figures show that the quality of the data is good
enough to provide positions with sub-arcsec accuracy, except for
$\ls$12\% of F band sources (i.e., 202 sources), and for $\sim$~13.5\%
of the entire source catalog (see Paper I for more details). These
small positional errors are the key to the high identification rate of
the C-COSMOS sources with optical and infrared counterparts (Civano et
al. 2009, Paper III).

\subsection{Count rates}

Vignetting corrected count rates for each source are obtained by
dividing the best-fit counts derived from {\it emldetect} for each
band and in each single field by the net exposure time, reduced by the
vignetting at the position of each source, as in the exposure
maps\footnote{http://hea-www.harvard.edu/$\sim$elvis/CCOSMOS.html}$^,$\footnote{http://irsa.ipac.caltech.edu/data/COSMOS/}. The
exposure maps are computed averaging over an area of 8 pixels to
smooth out CCD gaps and cosmetic defects, and are weighted with an
absorbed power-law spectral model with an energy index $\alpha_E =$0.4
and the Galactic column density of the COSMOS field,
N$_H$$=$2.7$\cdot$10$^{20}$cm$^{-2}$.

The errors on count rates at 68\% confidence level were then computed 
using the equation:
\begin{equation}
Error={\sqrt{ C_{s} + (1+a)\cdot B}\over{0.9 \cdot T_{expo}}}
\end{equation}
where C$_s$ are the source net counts estimated by {\it emldetect},
corrected to an area including 90\% of the PSF (see note 19), B are
the background counts from the {\it emldetect} background maps
(counts/pixel$^2$) multiplied by a circular area of radius
corresponding to $f_{psf}$$=$90\% and T$_{expo}$ is the vignetting
corrected exposure time at the position of the source from the
exposure maps. $a$ is a parameter which accounts for the fact that the
background at the source position is not known with infinite
precision. $a=1$ corresponds to the situation of a background area
equal to the source extraction area, which for {\it Chandra} is always
very small because of the very good PSF; $a=0$ would correspond to
assuming no uncertainty on the estimate of the average value of the
background. Unfortunately {\it emldetect} provides neither the B
errors, nor the information on the size of the region used to measure
the background counts. Because of the way {\it emldetect} estimates
the background counts, i.e. by a fit, using a sophisticated background
modeling (Cappelluti et al. 2007), we are in an intermediate situation
between the two extreme cases a$=$0 and a$=$1. For this reason, we
chose to adopt a$=$0.5.  This ensures that we are not under-estimating
the error on the background, even for sources close to problematic
areas like the edge of the field or CCD gaps. We chose an area
corresponding to $f_{psf}$$=$90\%, because this is the typical size of
the area where {\it emldetect} works for relatively bright sources.
We checked that the errors computed using Eq. 2 agree well with the
errors evaluated from aperture photometry (Sect. 6.4).

\begin{figure}
\includegraphics[width=8.5cm]{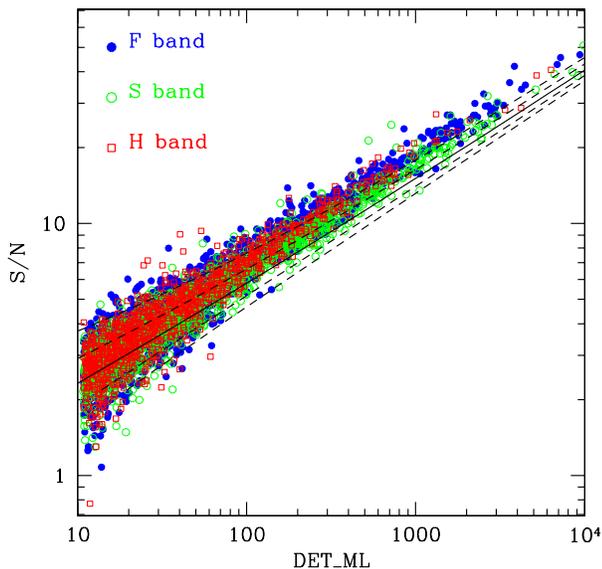}
\caption{The signal-to-noise ratio of each source as a function of
DET\_ML. Filled circles = F sources; open circles = S
sources; open squares = H sources.  The six diagonal black lines correspond
to the expectations assuming a background of 8, 4, 2, 1, and 0.5 counts
in the detection cell, from top to bottom.
\label{detml}}
\end{figure}

Fig. \ref{detml} plots the signal-to-noise ratio\footnote{Ratio
between the source count rate and the error on the source count rate
at 68\% confidence level} of each source as a function of
DET\_ML. Note the regular behavior of the signal-to-noise ratio,
which increases smoothly and monotonically with increasing DET\_ML, or
with decreasing P$_{random}$ (see Eq. 1), with a small dispersion
around the correlation. The six diagonal black lines show the
expectations computed for six values of the background in the
detection cell, from 0.5 counts to 8 counts. This range is centered on
$\sim$~4 counts, a value typical for the C-COSMOS survey (see Sect. 2),
and accounts for two effects: a) the differences in exposure time and
b) the difference sizes of the source extraction region as a function
of the off-axis angle, due to the variation of the {\it Chandra} PSF
with the off-axis angle. This range of background counts explains most
of the observed dispersion in Fig. \ref{detml}, especially for the
faintest sources. For the brightest sources in the F band, the best
fit DET\_ML is somewhat smaller than expected based on the signal-to-noise ratio,
even for the case of a background of 8 counts per detection cell. This
can be explained if the fit of bright sources is performed over an
area significantly larger than the 90\% f$_{psf}$ area, that so does
not fully optimize the signal-to-noise ratio. This shift is smaller
for the S band sources because of the smaller background in this band
with respect to the F band.

\subsection{Fluxes}

The {\it emldetect} count rates (R) were converted to fluxes (F$_x$)
using the formula: F$_x$$=$R/(CF$\cdot$10$^{11}$), where CF is the
energy conversion factor, that is evaluated by using spectra simulated
through {\it Xspec}\footnote{http://xspec.gsfc.nasa.gov/}, including
the appropriate on-axis response matrix and the chosen spectral
models. We used energy conversion factors of 0.742 counts erg$^{-1}$
cm$^{2}$, 1.837 counts erg$^{-1}$ cm$^{2}$, and 0.381 counts
erg$^{-1}$ cm$^{2}$ appropriate for a power-law spectrum with energy
index $\alpha_E=0.4$ and Galactic column density for the COSMOS field
(N$_H$$=$2.7$\cdot$10$^{20}$cm$^{-2}$), to convert the F count rate
into the 0.5-10 keV flux, the S count rate into the 0.5-2 keV flux, and the H
count rate into the 2-10 keV flux, respectively. We extend the F and H
bands up to 10 keV, to allow an easier comparison with the results of
literature. The conversion factors are sensitive to the spectral
shape: for $\alpha_E=1$ they change by $\sim~40\%$ in the F band, by
less than 5\% in the S band and by less than 25\% in the H band. For
absorbed power-law spectra with $N_H=10^{22}$ cm$^{-2}$ and
$\alpha_E=0.4$ or $\alpha_E=1.0$, the conversion factors change by up
to $\sim~46\%$ in the F band, by up to $\sim~17\%$ in the S band, and
by up to $\sim~18\%$ in the H band (see Tab. 3). The conversion factor
for the F band depends more strongly on the spectral shape because of
the wider band.

\begin{deluxetable}{lcccc}
\tabletypesize{\scriptsize} 
\tablecaption{Conversion factors for count rates to fluxes}
\tablewidth{0pt} 
\tablehead{ \colhead{$\alpha_E$ } & \colhead{N$_H$ } & \colhead{CF(F)\tablenotemark{a} } &
    \colhead{CF(S)\tablenotemark{a}} & \colhead{CF(H)\tablenotemark{a}} \\
    \colhead{ } & \colhead{ $10^{22}$ cm$^{-2}$} & \colhead{cts erg$^{-1}$ cm$^{2}$} & \colhead{ cts erg$^{-1}$ cm$^{2}$}
    & \colhead{cts erg$^{-1}$ cm$^{2}$} } 
\startdata
0.4 & 0.027  & 0.742 &  1.837 & 0.381\\
1   & 0.027  & 1.042 &  1.759 & 0.474\\
0.4 & 1      & 0.508 &  2.12 & 0.361\\
1   & 1      & 0.712 &  2.151 & 0.447\\
 \enddata 
\tablenotetext{a}{energy conversion factor to convert the F count rate into the 0.5-10 keV flux (CF(F)), the S count rate into the 0.5-2 keV flux (CS(S)), and the H count rate into the 2-10 keV flux (CF(H)) using the formula F$_x$$=$R/(CF$\cdot$10$^{11}$) and appropriate for a absorbed power-law spectra with the listed N$_H$ and $\alpha_E$.}
\label{tab2}
\end{deluxetable}

\subsection{Aperture photometry}

In addition to PSF fitting photometry, we have also performed standard
aperture photometry on the sources included in the final {\it
emldetect} catalog. We find an overall consistency between the two
estimates, with the {\it emldetect} count rates slightly larger, less
than 10 \%, than the count rates by aperture photometry.
For each source in the catalog, aperture photometry was performed in
F, S, and H band with the {\it yaxx}
tool\footnote{http://cxc.harvard.edu/contrib/yaxx/}. The aperture
photometry values are derived from event data for each individual {\it
Chandra} observation, where a source is located. Then for sources being
located in multiple observations, the aperture photometry is performed
in each of the multiple observations, and the corresponding multiple
aperture photometry values are combined to produce a single set of
values, using the appropriate method shown in Tab. \ref{merge}.

To extract source counts, circular regions of radii corresponding to
50\%, 90\%, and 95\% f$_{psf}$, centered on each source location, are
used for each observation, where the source is located. The radii are
calculated using the off-axis and azimuthal angles of the source in
each observation, and interpolating the circular f$_{psf}$ table
provided by the CXC calibration group to the nearest angles. Mean energies 2
keV, 1.2 keV, and 3.6 keV were chosen for the F, S, and H band,
respectively. To extract background counts, annuli with the inner edge
at the 95\% f$_{psf}$ radius plus 8 pixels, and with a width of 40
pixels are used. To limit contamination, all sources that overlap with
the source or background regions are masked by using circular
exclusion regions with the 95\% f$_{psf}$ radius. Exclusions can also
come from the CCD edge, with an 8 pixel padding inward from the
edge. Aperture fluxes for which the net source extraction area was less
than 75\% of the available area (i.e., the original circle prior to
exclusions) are not given in the catalog.

Using the region described above, photometry was extracted using the
CIAO tool {\it dmextract}. The source net counts were then corrected
for the fraction of f$_{psf}$. {\it dmextract} was also run on the
exposure maps with exactly the same regions in order to compute the
vignetting corrected exposure times, that are needed to compute the
source count rates. 

\begin{figure}
\includegraphics[width=8cm]{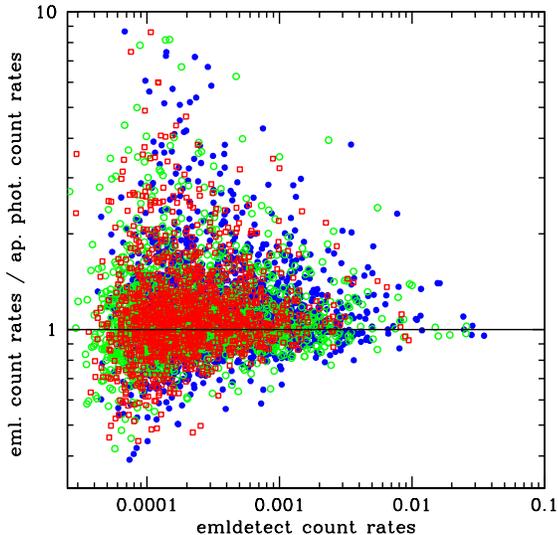}
\caption{ Ratio between the count rates evaluated by {\it emldetect}
and count rates evaluated by the aperture photometry as a function of
the {\it emldetect} count rates, for the F sources (blue filled dots),
S sources (green open dots), and H sources (red open squares). The
solid black line is the exact match between the {\it emldetect} count
rates and the aperture photometry count rates.
\label{emlap}}
\end{figure}

Fig. \ref{emlap} compares the count rates evaluated by {\it emldetect}
with the count rates evaluated by the aperture photometry. The median
and interquartile of the count rate ratios are 1.03$\pm$0.16,
1.08$\pm$0.19, 1.07$\pm$0.18 in the S, H, and F band,
respectively. 

\begin{deluxetable*}{lccc}
\tabletypesize{\scriptsize} 
\tablecaption{ Merge methods}
\tablewidth{0pt} 
\tablehead{ \colhead{Parameter } & \colhead{Symbol } & \colhead{Merge method } } \\
\startdata
exposure time corrected for the vignetting &  T$_{expo}$ & $\sum_i {T_{expo}}_i$$^1$  \\
counts &   T &  $\sum_i T_i$   \\
background counts & B &  $\sum_i B_i$    \\
net counts &  C$_s$ &   $\sum_i {C_s}_i$    \\
errors on counts & err$\_T$ & $\sqrt {\sum_i {{err\_ T}_i}^2}$ \\
errors on background counts & err\_$B$ & $\sqrt {\sum_i {{err\_ B}_i}^2}$ \\
errors on net counts & err$\_C_s$ & $\sqrt {\sum_i {{err\_ C_s}_i}^2}$ \\
count rates & R  &$\sum_i {{{R}_i}\cdot{{T_{expo}}_i}\over{{T_{expo}}} }$  \\
net count rates & R$_s$ &$\sum_i {{{R_s}_i}\cdot{{T_{expo}}_i}\over{{T_{expo}}} }$  \\
count rate errors &  err\_R &  $\sqrt{\sum_i {{{err\_R}_i}^2\cdot{{T_{expo}}_i}}\over{{T_{expo}}}}$ \\
net count rate errors &  err\_R$_s$ &  $\sqrt{\sum_i {{{err\_{R_s}}_i}^2\cdot{{T_{expo}}_i}}\over{{T_{expo}}}}$\\
\enddata 
\tablecomments{The index i indicates each of the observations where a source is located. }
\label{merge}
\end{deluxetable*}

\subsection{Upper limits}

If a source is not detected in one band, we give the 90\% upper limits
to the source count rates and fluxes in this band. The upper limits
are computed following as follows: if T is the total number of counts
measured at the position of a source not satisfying our detection
threshold, B are the expected background counts and X are the
unknown counts from the source, the 90\% upper limit on X (X(90\%))
can be defined as the number of counts X(90\%) that gives 10\%
probability to observe T (or less) counts. Applying the Poisson
probability distribution function, X(90\%) is therefore obtained by
iteratively solving for different X values the following equation:
\begin{equation}
0.1=e^{-(X+ B)}\sum_{i=0}^T { (X+ B)^i \over i!}
\end{equation}
(see e.g., Narsky 2000).
We collected the counts T both from a region of 5 arcsec radius and
from the aperture photometry discussed in Section 6.5. The results
were always statistically consistent with each other. The X(90\%)
upper limits derived with Eq. 3 do not take into account the
statistical fluctuations on the expected number of background
counts. In order to take the background fluctuations into
consideration, we used the following procedure: if $\sigma$(B) is the
root mean square of B (e.g., $\sigma$(B)$=$$\sqrt B$ for large B),
we estimated the 90\% lower limit on B as B(90\%) $=$B $-$ 1.282
$\cdot$ $\sigma$(B)\footnote{The value 1.282 is the value appropriate
for the 90\% probability (see e.g., Bevington P.R. and K. Robinson 1992).
 This approximate formula produces 90\% limits which differ by $\sim$~10\% (4\%) from the
exact estimate for values of B = 5 (10) in the extraction region,
corresponding to 0.064 (0.128) cts/arcsec$^2$ (see Fig. 3).}  and, as
a consequence, the ``correct" 90\% upper limit on X becomes:
\begin{equation}
       X_{corr}(90\%)\sim X(90\%) + 1.282 \cdot \sigma(B)                                
\end{equation}
We used X$_{corr}$(90\%) as upper limits for C-COSMOS sources. We
also evaluated the upper limits following the method described in
Kashyap et al. (2009). Comparing the upper limits obtained using the
two methods, we found that our upper limits are generally more
conservative (i.e., higher) than those which would be derived using
the method by Kashyap et al. (2009).

\section{Survey sensitivity and sky-coverage}

\subsection{Survey sensitivity}

In X-ray observations the sensitivity, i.e., the flux limit, is not
uniform in the field of view (FOV), due to two main reasons: (1) the
variable size of the PSF, that determines the background counts that
limit the source detection; and (2) the vignetting of effective area.
In C-COSMOS, where we have used multiple overlapping pointings giving
different PSFs and vignetting factors for each observation of each
source, the problem of assessing the sensitivity at each position in
the field of view is more complex than normal.  To evaluate the
C-COSMOS survey sensitivity, we have developed a dedicated procedure
by adapting the analytical method, used for the easier case of the
ELAS-S1 mosaic (Puccetti et al. 2006 and references therein), to the
more complicated C-COSMOS mosaic. In this procedure, the full C-COSMOS
field is divided into a grid of positions with spacing of 4 pixels,
i.e., 2 arcsec. This bin size is a suitable balance between the
spatial resolution in the C-COSMOS survey, and the ram memory required
for computing the sensitivity maps. At each point of the 2 arcsec
grid, we evaluated the minimum number of counts C$_{min}$ needed to
exceed the fluctuations of the background, assuming Poisson statistics
with a significance level equal to that used for the catalog (i.e.,
$2\cdot 10^{-5}$, see Sect. 6.1), according to the following formula:
\begin{equation}
P_{Poisson}=e^{-B}\sum_{k=C_{min}}^\infty{B^k \over k!}=2\cdot 10^{-5}
\end{equation}
where B is the total background counts computed at the position of
each point (P$_j$) of the grid by B$=$$\sum_{i=1}^n B_i$, where $i$ runs
from 1 to the number of overlapping fields at the position of each P$_j$
and B$_i$ are the background counts computed using the background map
of each {\it Chandra} pointing covering the position, in a
region centered at P$_j$ and of radius R$_i$. R$_i$ corresponds to a fixed
value of $f_{psf}$, and is evaluated from the distance of P$_j$
and the aim point of each single {\it Chandra} pointing covering the
position, using the CXC calibration.  We solved Eq. 5 iteratively to
calculate C$_{min}$. The count rate limit, R$_{lim}$, at each point of the grid is
then computed by:
\begin{equation}
R_{lim}= {C_{min}-B \over{f_{psf} \cdot T_{expo}}}
\end{equation}
where T$_{expo}$ is the total, vignetting corrected, exposure time at each
position of the grid, read from the merged C-COSMOS exposure
map (see notes 20, 22).

Finally, the flux limits at each P$_j$ are computed using the same
conversion factor used for the real C-COSMOS sources.  This procedure
is applied to the S, H, and F bands to produce binned sensitivity
maps.

\subsection{Sky-coverage}

The ``sky-coverage'' is the integral of the survey area covered down
to a given flux limit, as a function of the flux in the sensitivity
map. The solid lines in Fig. \ref{complet2} are the normalized
sky-coverages, computed using the procedure described above and
adopting $f_{psf}=0.5$. We studied how the sensitivity maps and the
sky-coverage depend on the assumption on $f_{psf}$ and found that they
change less than few per cent for $f_{psf}$ values up to 0.90. We also
studied how the sensitivity maps change using different $f_{psf}$
values at different off-axis angles where a single source is observed,
finding again very little change. The reason for this behaviour is the
relatively low backgound within each R$_i$, even at large off-axis
angles.

\begin{figure}
\includegraphics[width=9cm]{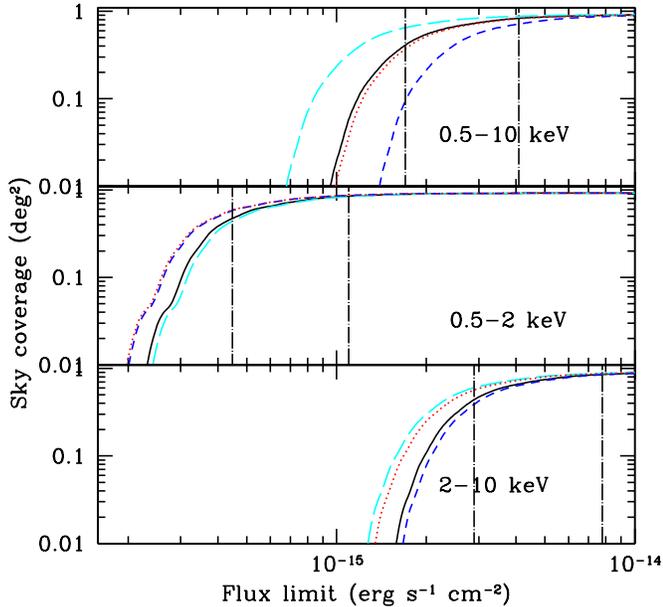}
\caption{The sky-coverage calculated as in Sect. 7.2 for the 0.5-10
keV ({\it top panel}), 0.5-2 keV ({\it middle panel}), and 2-10 keV
({\it bottom panel}) band. The {\em black solid} lines represent the
sky-coverages evaluated with the baseline model (i.e., power-law
spectra with $\alpha_E=0.4$ absorbed by Galactic N$_H=2.7 \cdot
10^{20}$ cm$^{-2}$). The {\em cyan long-dashed} lines represent the
sky-coverages for power-law spectra with $\alpha_E=1$ absorbed by
Galactic N$_H=2.7 \cdot 10^{20}$ cm$^{-2}$. The {\em blue
short-dashed} lines represent the sky-coverages for power-law spectra
with $\alpha_E=0.4$ absorbed by N$_H= 10^{22}$ cm$^{-2}$. The {\em red
dotted} lines represent the sky-coverages for power-law spectra with
$\alpha_E=1$ absorbed by N$_H= 10^{22}$ cm$^{-2}$. The {\em black
dot-long dashed} vertical lines represent the fluxes correspondig to
the 90\% and 50\% completeness, respectively.
\label{skycc}}
\end{figure}

A relatively large uncertainty in the sensitivity maps and sky
coverage computation is instead, the unknown spectrum of the sources
near the detection limit. The magnitude of this uncertainty depends on
the width of the energy band, and therefore is largest in the F band.
To estimate the magnitude of the uncertainty, we calculated the sky
coverage for power-law spectra with $\alpha_E=1.0$, and for absorbed
power-law spectra with $\alpha_E=0.4$ or $\alpha_E=1.0$ and
N$_H=10^{22}$ cm$^{-2}$, in addition to the baseline case
($\alpha_E=0.4$, N$_H=2.7 \cdot 10^{20}$ cm$^{-2}$; see
Fig. \ref{skycc}). At the flux limits corresponding to 90\%
completeness (see Tab. 2) the deviations are less than 3\%,
$\sim$~3\%, and $\sim$~16\% for the S, H, and F bands,
respectively. This uncertainty related to the unknown spectrum of the
sources becomes significant only at fluxes below the 50\%
completeness.

\subsection{The $\log N$ -- $\log S$}

\begin{figure*}
\begin{tabular}{ccc}
\includegraphics[width=6cm]{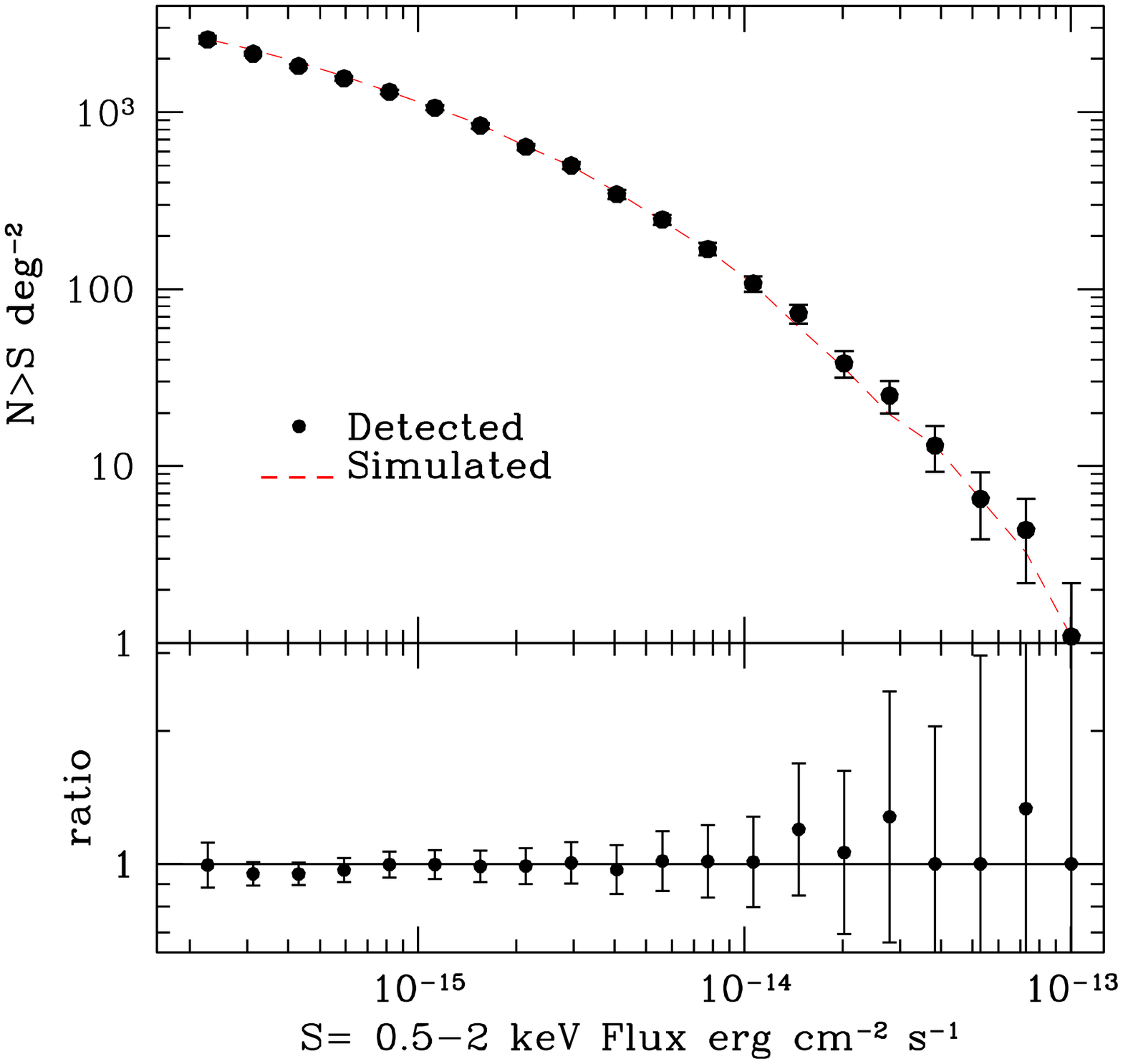}
\includegraphics[width=6cm]{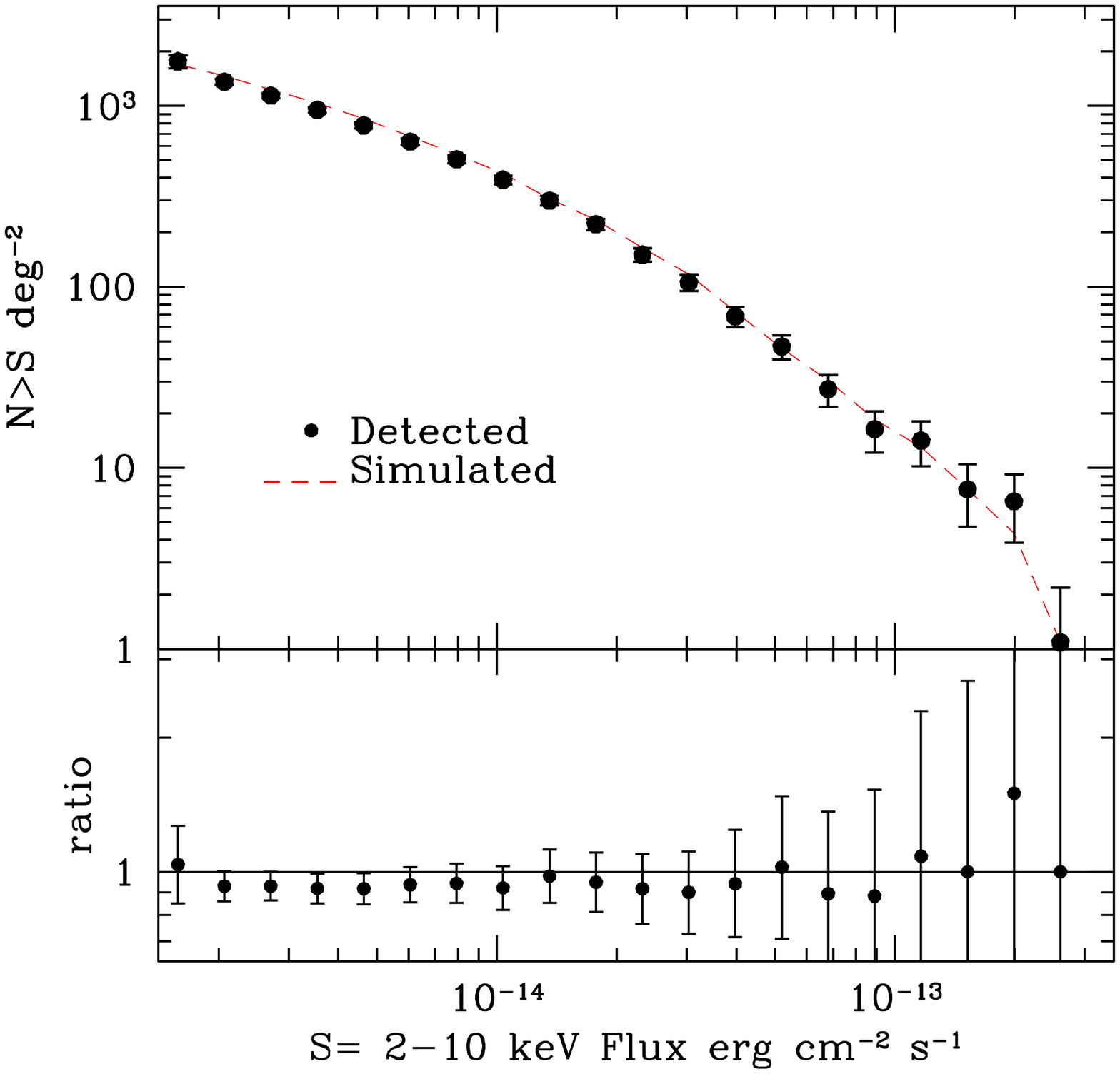}
\includegraphics[width=6cm]{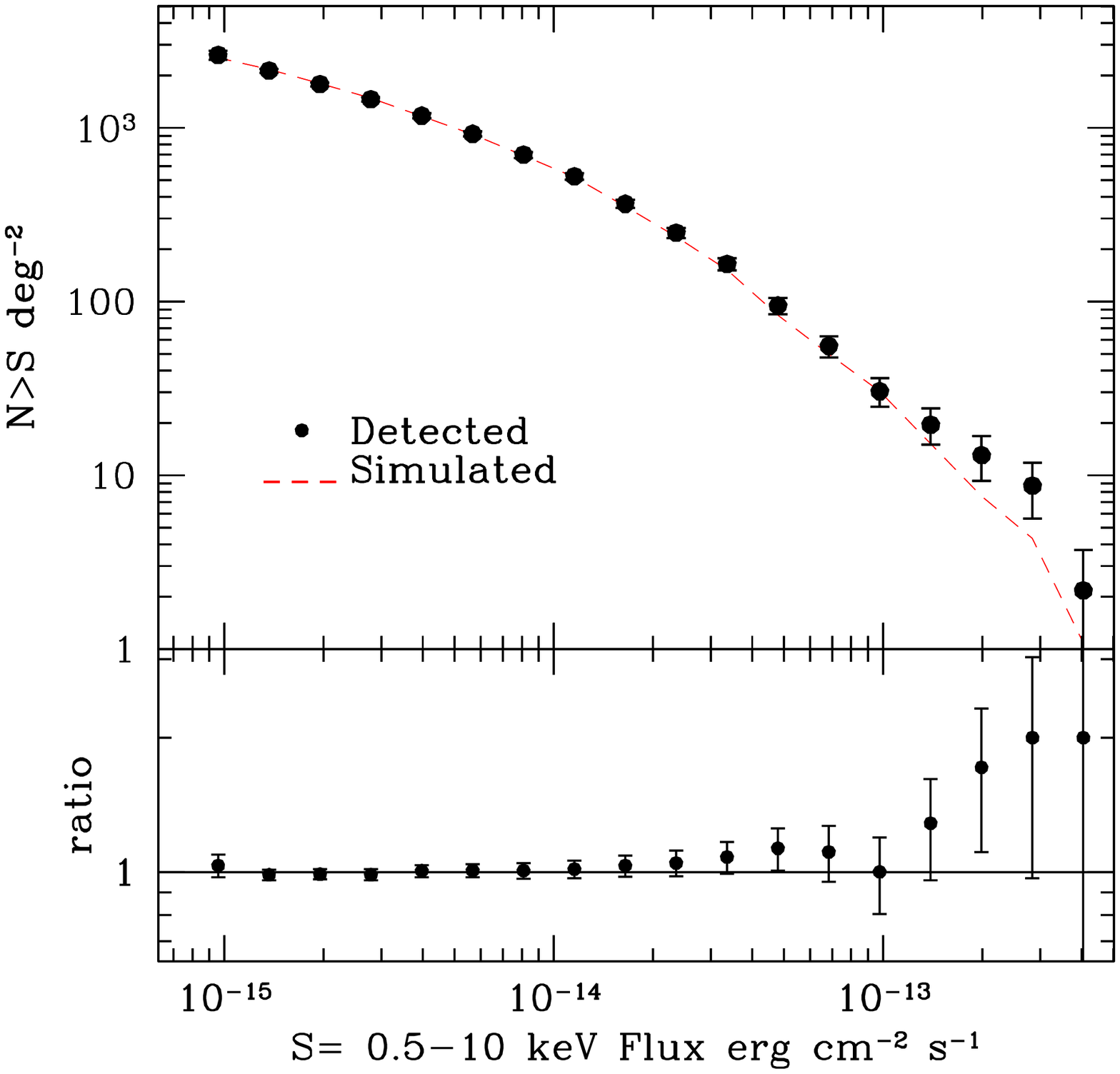}
\end{tabular}
\caption{$\log N$ -- $\log S$ curves of the simulated sources (dashed
red curves) compared with the $\log N$ -- $\log S$ curves of the sources
detected in the simulations (black dots): {\it top left panel}: S
band, {\it top middle panel}: 2-10 keV band, {\it top right panel}: 0.5-10 keV band. Ratio between $\log N$ -- $\log S$ curves of the simulated sources and the $\log N$ -- $\log S$ curves of the sources
detected in the simulations: {\it bottom left panel}: 0.5-2 keV
band, {\it bottom middle panel}: 2-10 keV band, {\it bottom right panel}: 0.5-10 keV band. 
\label{lnlssim}}
\end{figure*}

We used the catalogs of the sources detected in the simulations in the
S, H, and F bands, and the sky-coverage curves computed in Sect. 7.2 to
obtain the number counts ($\log N$ -- $\log S$) of the sources
detected in the simulations. We cut the catalogs in the S, H, and F
band at a signal-to-noise ratio higher than 2, 2.5, and 2.8, respectively.  These
cuts are introduced because: (1) we do not correct for Eddington bias,
which may be strong (up to 30-50\%) at the lowest flux limits; (2) low
signal-to-noise implies a large statistical uncertainty in the flux,
which in turn would introduce a large uncertainty on the number counts
at the lowest fluxes; (3) at the lowest fluxes, the sky-coverage is
small, and the relative statistical and systematic errors are
therefore large, again introducing large uncertainties in the number
counts.  We chose the signal-to-noise thresholds by requiring that the
deviations between the $\log N$ -- $\log S$ computed from the detections
and the input $\log N$ -- $\log S$ are smaller than 5\%.  The $\log
N$-$\log S$ are shown in Fig. \ref{lnlssim}. The flux limits implied
by the signal-to-noise thresholds are $\sim~2.3\cdot 10^{-16}$,
$\sim~1.6\cdot 10^{-15}$, and $\sim~9.6\cdot 10^{-16}$ erg s$^{-1}$
cm$^{-2}$ for the 0.5-2 keV, 2-10 keV, and 0.5-10 keV band, respectively. These
flux limits are fully consistent with the flux limits of the $\log N$
-- $\log S$ derived from the observed data (see Paper I).

\section{Comparison with AEGIS-X}

\begin{figure*}
\includegraphics[width=8cm]{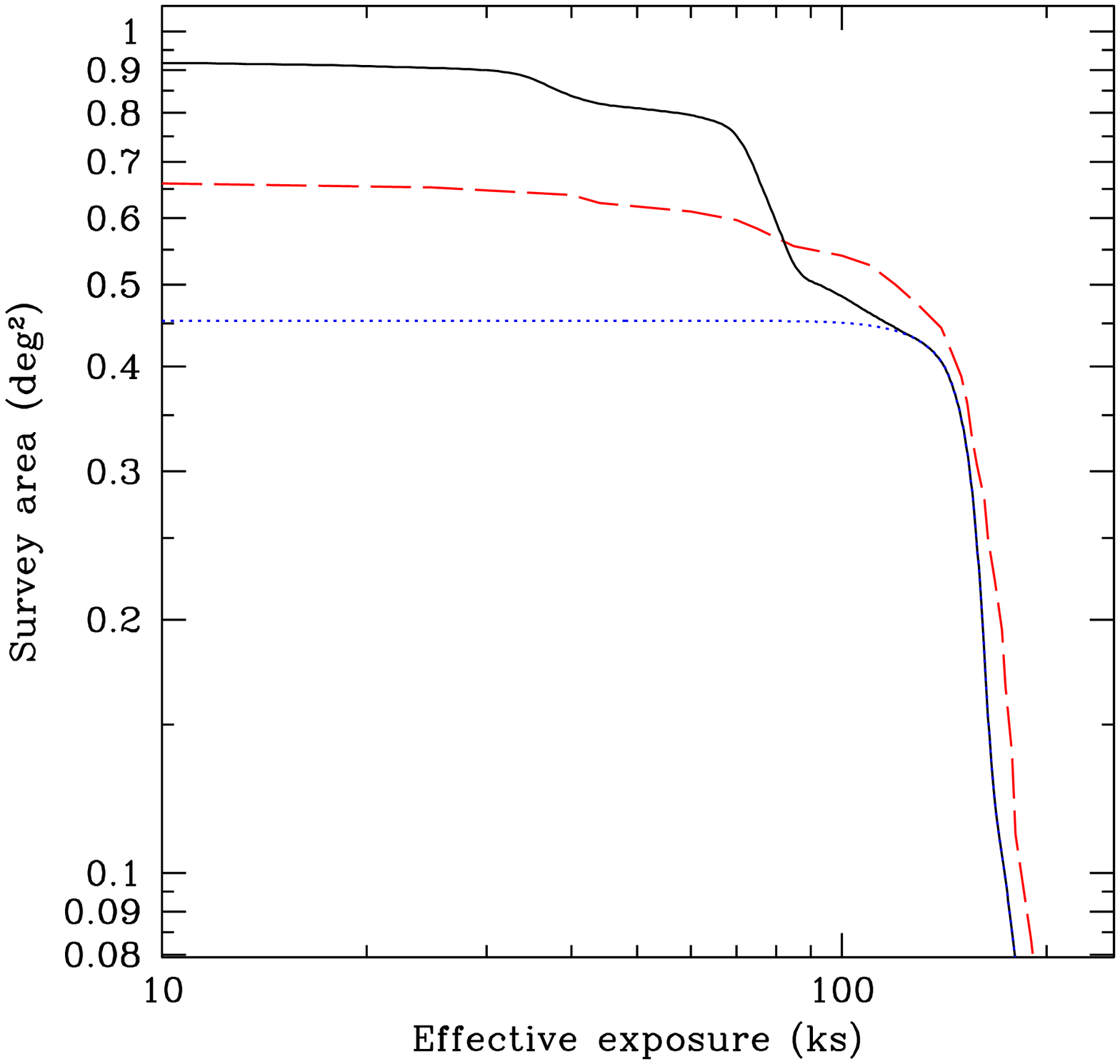}
\includegraphics[width=8cm]{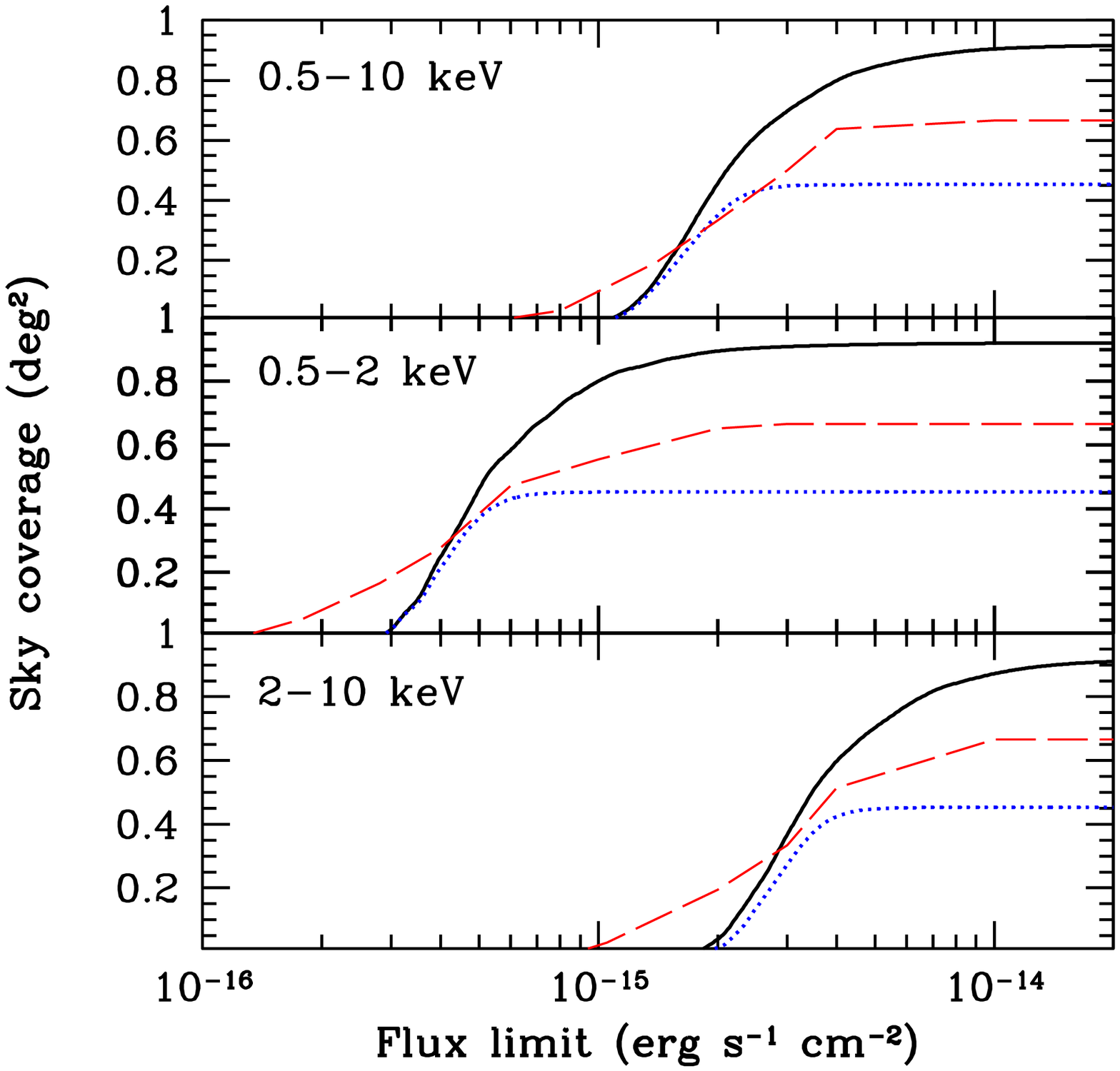}
\caption{{\it Left panel:} the survey area coverage as a function of the
effective (i.e., vignetting corrected) exposure time. {\it Right
panel:} the sky-coverage calculated as in Sect. 7.2, using a
significance level of $4\cdot 10^{-6}$, in the 0.5-10 keV band
({\it upper panel}), in the 0.5-2 keV band ({\it middle panel}), and in the
2-10 keV band ({\it bottom panel}). In both panels, the solid black lines represent
the total C-COSMOS survey, the dotted blue lines represent the central
C-COSMOS $\sim$~0.45 deg$^2$ area with the best exposure and the red
dashed lines represent the AEGIS-X survey.
\label{AEGIXcc}}
\end{figure*}

{\it Chandra} was used to perform a survey somewhat similar to
C-COSMOS in the Extended Groth-Streep (AEGIS-X, Laird et
al. 2009). The 1.6 Ms AEGIS-X survey is made of 8 ACIS-I pointings,
each of a nominal 200 ksec exposure, with very little overlap,
covering $\sim$~0.67 deg$^2$.  While the effective exposure time and
area coverage are similar to C-COSMOS (see Fig. \ref{AEGIXcc}), the
tiling is completely different. In C-COSMOS each source in the central
area is observed at four to six different off-axis angles, while in
AEGIS-X each source is observed only at one off-axis angle.

To compare the two surveys quantitatively, we cut the C-COSMOS catalog
at the same significance level used for AEGIS-X (i.e., $4\cdot
10^{-6}$ or DET\_ML=12.4, Laird et al. 2009). We also recomputed the
C-COSMOS sky-coverage using the same significance level. Fig.
\ref{AEGIXcc} compares the C-COSMOS sky-coverage to the AEGIS-X one
computed without the Bayesian correction for the Eddington bias.  The
C-COSMOS sky-coverage has a significantly sharper drop toward lower
fluxes than the AEGIS-X sky-coverage.  This means that the sensitivity
in C-COSMOS is more uniform over the field than in AEGIS-X, while the
AEGIS-X tiling reaches fainter limiting fluxes than C-COSMOS. The
estimated AEGIS-X flux limit in the S band is 50\% deeper than
C-COSMOS, while the flux limits in the H and F bands are about twice
as deep as C-COSMOS, albeit in small areas. The deeper AEGIS-X flux
limit in the H and F bands with respect to the S band depends on the
higher internal background in these bands and on the smaller typical
source extraction regions in the areas of best sensitivity of AEGIS-X
with respect to C-COSMOS. In fact, AEGIS-X has a PSF better than $\sim
~1$ arcsec over an area of $\sim$~0.15 deg$^2$, while the complex
C-COSMOS tiling implies effective source extraction regions of radii
of $\sim$~3 arcsec over most of the area.

The more characteristic flux limits
corresponding to 90\% completeness in the F, S, and H bands are similar
in C-COSMOS and AEGIS-X, while the AEGIS-X flux limits corresponding to
50\% completeness in the F, S, and H bands are lower than C-COSMOS by a factor 2-3 (see Tab. \ref{aeg}).

\begin{deluxetable*}{lccccccc}
\tabletypesize{\scriptsize} 
\tablecaption{ Comparison between C-COSMOS and AEGIS-X}
\tablewidth{0pt} 
\tablehead{ \colhead{Parameter } & \colhead{units } & \colhead{C-COSMOS F  } & \colhead{AEGIS-X F  } & \colhead{C-COSMOS S  } &
    \colhead{AEGIS-X S }& \colhead{C-COSMOS H  } & \colhead{AEGIS-X H  } \\
    }
\startdata
90\% Completeness\tablenotemark{a} & \cgs & $4.3\cdot10^{-15}$ & $4.0\cdot10^{-15}$ & $1.1\cdot10^{-15}$ & $1.1\cdot10^{-15}$& $8.0\cdot10^{-15}$& $6.2\cdot10^{-15}$ \\
50\% Completeness\tablenotemark{a}& \cgs & $1.8\cdot10^{-15}$ & $6\cdot10^{-16}$   & $5.1\cdot10^{-16}$ & $1.4\cdot10^{-16}$ & $1.8\cdot10^{-15}$ & $9\cdot10^{-16}$ \\ 
\hline
observed source densities\tablenotemark{b}&sources$\cdot$deg$^{-2}$& 2110$\pm$68  & 1830$\pm$52  & 1700$\pm$61 & 1550$\pm$40 & 1320$\pm$54  & 1110$\pm$41 \\
\hline
number of spurious sources\tablenotemark{c} & sources &5& 5 & 4 &5 &3 &5\\
\hline
number of sources\tablenotemark{d} & sources &1655 &  1221 & 1340 & 1032&1017 & 741\\
 \enddata 
\tablenotetext{a}{Completenesses are evaluated using a significance
level of $4\cdot10^{-6}$.}

\tablenotetext{b}{The observed source densities are evaluated in the total AEGIS-X
area, and in the central $\sim$~0.45 deg$^2$ area in the C-COSMOS,
which has similar effective exposure of the AEGIS-X survey, using a
significance level of $4\cdot10^{-6}$.}

\tablenotetext{c}{In C-COSMOS the spurious sources are evaluated using
a significance level of $2\cdot 10^{-5}$ in the total field, for each
band. For AEGIS-X Laird et al. (2009), using simulations, found 0.58
spurious sources per 200 ksec field per band using a significance
level of $4\cdot10^{-6}$, corresponding to 5 spurious sources in
the full AEGIS-X survey, each band.}

\tablenotetext{d}{Number of sources detected in each band in the total
fields, using a significance level of $2\cdot 10^{-5}$ and
$4\cdot10^{-6}$ for C-COSMOS and AEGIS-X, respectively.}

\label{aeg}
\end{deluxetable*}

The more uniform sensitivity of C-COSMOS over the field reaches a higher source density (see Tab. \ref{aeg}).

C-COSMOS we estimate a slightly lower number of spurious sources at a higher
significance level (i.e., $2\cdot 10^{-5}$ vs. $4\cdot 10^{-6}$, see
Tab. \ref{aeg}), with respect to AEGIS-X survey. The number of
spurious sources is roughly given by the product of the significance
level times the number of independent detection cells in the field.
The combination of different PSFs at each C-COSMOS position produces
an effective source extraction region of $\sim$~3 arcsec radius, i.e.,
significantly wider than the Chandra PSF at off-axis angles smaller
than 5-6 arcmin. This means that the number of independent cells per
unit area is smaller in C-COSMOS than in AEGIS-X.  In conclusion, the
lower number of spurious detections in C-COSMOS with respect to
AEGIS-X at a given significance level is due to the fact that each
field is observed more than once at different off-axis angles and
therefore with different PSFs.

\section{Conclusion }

The complex tiling of C-COSMOS survey required the development of a
tailored multistep procedure to fully exploit the data. Detailed
simulations were used to test different detection (sliding cell and
wavelet) and photometry (PSF fitting and aperture photometry)
algorithms. In particular, we compared the results obtained using the
SAS {\it eboxdetect} and {\it emldetect} tasks, used for the
XMM-COSMOS survey (Cappelluti et al. 2007, 2009), with those obtained
using the {\it PWDetect} code (Damiani et al. 1997).  Through these
tests we selected a procedure consisting in first identifying source
candidates using {\it PWDetect}, and then performing accurate PSF
fitting photometry and evaluating aperture photometry for each source
candidate.  In this way we obtained subarcsec source localizations and
accurate photometry even for partly blended sources.

We set a threshold for source detection to $P=2\cdot 10^{-5}$, which
implies a completeness of 87.5\% and 68\% for sources with at least 12
and 7 F band counts, respectively, and 3 to 5 spurious detections in
the F band at the same count limits, respectively.

We evaluated the survey sensitivity and the sky-coverage, through an
analytical method, tuned using simulations.  We then evaluated the
$\log N$ -- $\log S$ of the detected sources in the simulations down
to F, S, and H band flux limits of F$_{x}$$\sim~2.3\cdot 10^{-16}$,
$\sim~1.6\cdot 10^{-15}$, and $\sim~9.6\cdot 10^{-16}$ erg s$^{-1}$
cm$^{-2}$, respectively.

Finally we compared the C-COSMOS survey to the AEGIS-X survey, a {\it
Chandra} survey with similar sky-coverage and total exposure time, but
using non overlapping ACIS-I pointings.  We found that the complex
tiling of C-COSMOS helps in obtaining a contiguous area with uniform
sensitivity and somewhat higher source density. The overlap of several
pointings with different PSF at the same position produces an
effective source extraction region of $\sim$~3 arcsec radius,
i.e., significantly wider than the Chandra PSF at off-axis angles
smaller than 5-6 arcmin. This produces a number of independent
detection cells per unit area smaller than in a single ACIS-I pointing
survey like AEGIS-X, which in turn implies a smaller number of
spurious sources at each given detection threshold.

\acknowledgments

This research has made use of data obtained from the {\em Chandra}
Data Archive and software provided by the {\em Chandra X-ray Center}
(CXC) in the application packages CIAO and Sherpa.

This work was supported in part by NASA {\em Chandra} grant number
GO7-8136A (Martin Elvis, SAO, Francesca Civano, Marcella Brusa, Alexis
Figonuenov), NASA contract NAS8-39073 (Chandra X-ray Center), and by
NASA/ADP grant NNX07AT02G.  In Italy this work is supported by
ASI/INAF contracts I/023/05/0, I/024/05/0, I/026/07/0, I/088/06, and
2007/1.06.10.08, by PRIN/MUR grant 2006-02-5203. In Germany this
project is supported by the Bundesministerium f\"{u}r Bildung und
Forschung/Deutsches Zentrum f\"{u}r Luft und Raumfahrt and the Max
Planck Society. In Mexico, this work is supported by CONACyT 83564 and
PAPIIT IN110209.

\end{document}